\documentclass[useAMS,usenatbib]{mn2e}
\usepackage{epsfig}

\def\simgt{\lower.5ex\hbox{$\; \buildrel > \over \sim \;$}}
\def\simlt{\lower.5ex\hbox{$\; \buildrel < \over \sim \;$}}

\title[Outflows in star forming dwarf
  galaxies]{Galactic fountains and outflows in star forming dwarf
  galaxies: ISM expulsion and chemical enrichment}
 
\author[C. Melioli, F. Brighenti, A. D'Ercole]
{C. Melioli,$^{1,2}$\thanks{E-mail:
cmelioli@astro.iag.usp.br; fabrizio.brighenti@unibo.it; 
annibale.dercole@oabo.inaf.it} 
F. Brighenti$^{3}$ and A.D'Ercole$^{1}$ \\
$^{1}$INAF-Osservatorio Astronomico di Bologna, via Ranzani 1, 40126 Bologna, 
Italy\\
$^{2}$IAG - Universidade de S\~ao Paulo, Rua do Mat\~ao 1226, S\~ao Paulo, Brazil \\
$^{3}$Dipartimento di Fisica e Astronomia, Universit\`a di Bologna, via Ranzani 1, 
40126 Bologna, Italy}

\begin{document}

\date{Accepted ... Received ...; in original form ...}

\pagerange{\pageref{firstpage}--\pageref{lastpage}} \pubyear{2014}

\maketitle

\label{firstpage}

\begin{abstract}
We investigated the impact of supernova feedback in gas-rich dwarf
galaxies experiencing a low-to-moderate star formation rate, typical
of relatively quiescent phases between starbursts. We calculated the
long term evolution of the ISM and the metal-rich SN ejecta using 3D
hydrodynamic simulations, in which the feedback energy is deposited
by SNeII exploding in distinct OB associations. We found that a
circulation flow similar to galactic fountains is generally
established, with some ISM lifted at heights of one to few kpc above
the galactic plane. This gas forms an extra-planar layer, which falls
back to the plane in $\approx 10^8$ yr, once the star formation stops.
Very little or no ISM is expelled outside the galaxy system for the
considered SFRs, even though in the most powerful model the SN 
energy is comparable to the gas binding energy.
The metal-rich SN ejecta is instead more vulnerable to the feedback
and we found that a significant fraction ($25-80$\%) is vented in the
intergalactic medium, even for low SN rate ($7\times 10^{-5} - 
7\times 10^{-4}$ yr$^{-1}$). About half of the metals
retained by the galaxy are located far ($z >$ 500 pc) from the galactic
plane. Moreover, our models indicate that the circulation of the
metal-rich gas out from and back to the galactic disk is not able to
erase the chemical gradients imprinted by the (centrally concentrated)
SN explosions.

\end{abstract}


\section{Introduction} 
 \label{sec:introduction}

 Dwarf galaxies evolution is likely affected by both external and
 internal processes such as tidal and ram pressure stripping, and galactic winds
 \citep[see][for a detailed review]{tolstoy09}.  In particular, the
 interplay between Type II supernova (hereafter indicated simply as
 SN) explosions, stellar winds and interstellar medium (ISM) has
 important effects on the formation and the dynamical and chemical
 evolution of these systems.  Main characteristics of a dwarf galaxy
 are low metallicities, low surface brightness and low mass. Given
 their shallow potential, these galaxies can in principle lose a
 significant fraction of their ISM during the burst of SN explosions
 \citep[e.g.][]{dekel86, deyoung94, maclow99, dercole99}, and this
 could lead to a cessation of star formation and to the inability to
 retain newly synthesized metals \citep[external processes as ram
 pressure and tidal stripping are also expected to play a role; see,
 e.g., ][]{mayer06, marcolini06}.  The SN feedback may be the key to
 understand the possible link between the evolution of dwarf irregular
 galaxies (dIrrs) to dwarf spheroidal galaxies (dSphs), as pointed out
 by many authors \citep[see, e.g.,][]{ loose86, dekel86, silk87,
   davies88}.  However, while it is plausible that the star formation
 feedback has a crucial impact in the life of dwarf galaxies, the
 complexity of the interaction between the supernovae and the ISM
 still prevents a detailed comprehension of this process.

 The effect of the star formation heating must depend, among other
 things, on the star formation rate (SFR), the ISM density and spatial
 distribution and the depth of the galactic potential well.  A large
 fraction of (irregular and blue compact) dwarf galaxies host star
 formation (SF) episodes, and clustered SNe are responsible for
 the formation of superbubbles, outflows or galactic winds. However,
 without a direct measurement of the gas dynamics, it is difficult
 to understand whether the outflowing gas can escape the gravitational
 potential of the dwarf galaxy or it remains bound to the system to
 fall back at later times \citep[cf.][and references therein]{creasey13}.

   Recently, a number of high resolution numerical models of
   feedback from supernovae have been worked out to study in detail
   the role of the magnetic field \citep{hill12} or to obtain
   physically motivated sub-grid description of winds that can be
   implemented in cosmological hydrodynamic simulations and
   phenomenological models \citep{creasey13}. These authors, as others
   before
   \citep{korpi99,deavillez00,deavillez01,deavillez04,deavillez05,deavillez05b}
   simulate only a small volume of galactic disk in order to obtain high
   spatial resolutions. SPH
   simulations with pc-scale resolutions taking into account the whole
   galaxy have been presented by \citet{hopkins12}.

 In this paper we study the large scale and long term evolution of the gas
 ejected by SN explosions in a typical dwarf galaxy. Our aim is to
study the relation between SFR, mass loss and metallicity
 gradient evolution.  Abundance gradients in dwarf galaxies are
 observed to be essentially flat \citep[e.g.][and references
 therein]{lagos09,werk11,lagos12}, leading to the conclusion that one
 or more physical processes exist able to distribute metals throughout
 the gas disk.  Metal mixing may be favoured by SN driven winds
 carrying metals which can fall back down at large radii, by gas disk
 turbulence and by past interactions which have led to the accretion of
 external gas.  In this work we
 test the impact of gas flows (wind, outflow, fountains) on the
 metal abundance evolution.

Outflows and galactic winds were studied by many authors both
 analytically \citep[see, e.g.,][]{shapiro76, bregman80} and
 numerically by hydrodynamical simulations \citep[e.g.][]{ tomisaka86,
   dercole99, maclow99, strickland00, recchi01, mori02, recchi02,
   fragile04, hopkins12, recchi13, melioli13}. 
In these works it is investigated how the development
and evolution of galactic winds depend on the gravitational potential,
total gas and stellar mass, ISM density and temperature, SFR.

\citet{melioli08, melioli09} considered the occurrence
of galactic fountains in Milky Way-like galaxies. They showed that the
 majority of the gas lifted up in our Galaxy by the Galactic fountains
 falls back on the disk remaining quite close to the place where the
 fountains originated, thus not influencing the radial chemical
 gradients on large scale. 
However, in the case of dwarf galaxies,
it is plausible that the smaller size and the shallower potential
allow for a more effective redistribution of metals incorporated
in galactic flows.

To test such a possibility, we run 3D numerical simulations of star
formation feedback in a galaxy of total mass of $\sim 10^{10}$ M$_\odot$, taking
 into account different SFRs and different durations of the SN
 explosions.  The numerical grid
 used here includes the whole rotating galaxy, and the adaptive mesh
 scheme allows us to follow in detail the ejected gas, while the
 remaining galactic volume is mapped at a lower resolution.

 In the following sections, we describe the numerical procedure (Section 2).
 The basic results of the simulations are outlined in Section 3,
 while in Section 4 we discuss in detail the evolution of
 each model. Finally, in Section 5 we draw our conclusions.

\section{The models}
\label{sec:models}

\subsection{The dwarf galaxy}
\label{subsec:dwarf}

 In this Section we describe the dwarf galaxy model. We set the
  initial conditions for the ISM following the procedure outlined in
  \citet{melioli08}. In brief, we first assume a mass model for the
  galaxy, which includes the contribution of a stellar disk and of a
  spherical dark matter halo. Then, we build the ISM in rotational
  equilibrium in the galactic potential, by specifying the 3D density
  distribution and computing the temperature and azimuthal velocity
  distributions.

  The gravitational potential of the stellar disk is assumed to be
  generated by a density distribution following a flattened King
  profile
\begin{equation}
\label{eq:rhodisk}
\rho_{\star}(r)={\rho_{\star,0} \over \left [1+\left (R/
      R_{\star,{\rm c}}\right)^2 +\left (z/z_{\star,{\rm c}}\right)^2\right]^{3/2}}.
\end{equation}
\noindent
where $\rho_{\star,0}$ is the central density of the stars, and
$R_{\star,{\rm c}}$ and $z_{\star,{\rm c}}$ are the core radii.  In
this model we take $\rho_{\star,0}=2 \times 10^{-24}$ g cm$^{-3}$,
$R_{\star,{\rm c}}$ = 1.2 kpc and $z_{\star,{\rm c}}$ = 0.67 kpc.  The
stellar distribution is assumed to extend up to $R_{\star,{\rm t}}=4.8$
kpc and $z_{\star,{\rm t}}=2.6$ kpc. The gravitational potential $\Phi
_{\star}$ associated to this stellar disk is computed numerically
following the method described in \citet{brighenti96}.

\begin{figure}
\begin{center}
\psfig{figure=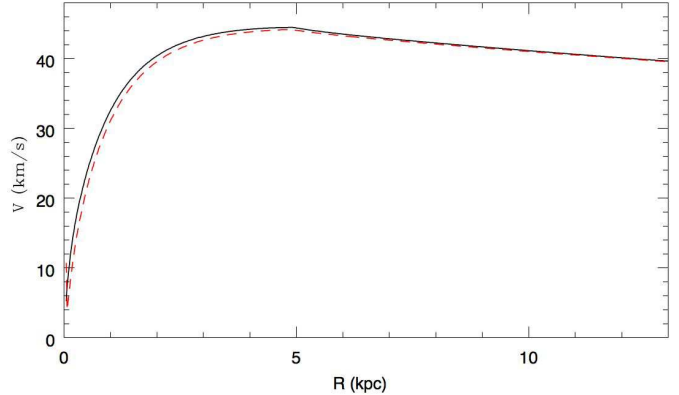,width=0.5\textwidth}
\end{center}
\caption{Circular velocity profile (solid line) and ISM rotational
  velocity profile in the $z=0$ plane.
}
\label{fig:star}
\end{figure}

Finally, the gravitational potential of the dark matter halo is assumed to 
follow the Navarro, Frenk and White profile \citep{navarro96}
\begin{equation}
\label{eq:phiblg}
\Phi_{\rm dm} (r)=-{{G M_{\rm vir}}\over{r_{\rm s}f(c)}}{\ln(1+x) 
\over x},
\end{equation}
\noindent
where M$_{\rm vir}$ is the mass at the virial radius $r_{\rm vir}$
(defined as the radius where the average density is $102 \rho_{\rm
  crit}$, being $\rho_{\rm crit}$ the cosmic critical density),
$x=r/r_{\rm s}$, $r$ is the spherical radius, $r_{\rm s}$ is a scale
radius, $c$ = $r_{\rm vir}/r_{\rm s}$ is the concentration, and $f(c)$
= ln$(1+c)-c/(1+c)$.  In this model we have M$_{\rm vir}$ = 10$^{10}$
M$_{\odot}$, $r_{\rm vir}$ = 55.3 kpc, $r_{\rm s}$ = 2.94 kpc and $c =
18.8$, adopting a $\Lambda$ cold dark matter cosmological universe
with $\Omega_{\rm M}$=0.27, $\Omega_{\Lambda}$=0.73 and $H_0$=71 km
s$^{-1}$ Mpc$^{-1}$.  The resulting circular velocity as a function of
radius (in the $z=0$ plane) is shown in Figure \ref{fig:star} (solid
line).

\begin{figure}
\begin{center}
\psfig{figure=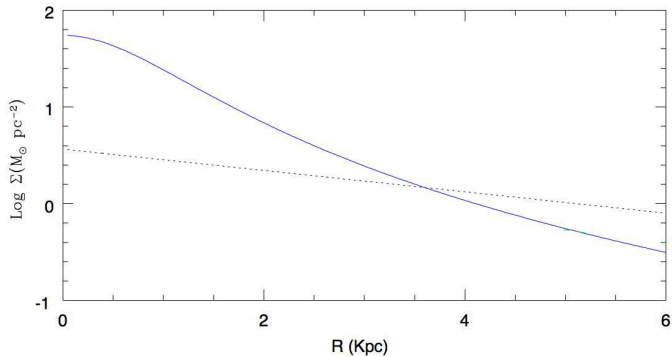,width=0.5\textwidth}
\end{center}
\caption{
Stellar (solid line) and gas (dotted line) mass surface density 
for the dwarf galaxy model (face-on). Distances are given in kpc and densities are 
in M$_{\odot}$/pc$^2$. The values refer only to the
  ``upper half galaxy'' ($z>0$) covered by the computational grid. The values for the whole
galaxy are just twice as those indicated.
}
\label{fig:sdgal}
\end{figure}

The assumed 3D gas density distribution writes:

\begin{equation}
\label{eq:ism}
\rho = {\Sigma_{\rm d} \over 2z_{\rm d}}\exp\left(-{R\over R_{\rm d}}-
{|z| \over z_{\rm d}}\right ),
\end{equation}

\noindent
where $R$ is the cylindrical radius, $z$ is the vertical height,
$R_{\rm d}$ and $z_{\rm d}$ are respectively the radial and the
vertical scale height of the disk, and $\Sigma_{\rm d}$ is
a characteristic surface density. In this study we assume $R_{\rm d} =
4$ kpc, $z_{\rm d} = 170$ pc and $\Sigma_{\rm d} =8 $ M$_{\odot}$
pc$^{-2}$.  The total gas mass within the computational box (see below) is
$M_{\rm g} \sim 1.5 \times 10^8$ M$_{\odot}$.   While these
  parameters are not designed to describe any specific galaxy, they
  are generally consistent with observed irregular dwarf galaxies,
  and, in particular, with the low surface brightness dwarf galaxies
  studied by \citet{vanzee97}. The stellar and gas surface densities
are shown in Figure \ref{fig:sdgal}, while the values of the galactic
parameters are summarized in Table \ref{tab:gal}.

\begin{table*}
 \centering
 \begin{minipage}{140mm}
  \caption{Galactic parameters.}
  \label{tab:gal}
  \begin{tabular}{@{}ccccccccccccc@{}}
  \hline
 $M_{\rm g}$ & $R_{\rm d}$ & $z_{\rm d}$ & $\Sigma_{\rm d}$ & 
$M_{\rm vir}$ & $r_{\rm vir}$ & $r_{\rm s}$ 
& $c_{102}$ & $M_{\star}$ & $R_{\star,{\rm t}}$ 
& $R_{\star,{\rm c}}$ & $z_{\star,{\rm t}}$ & $z_{\star,{\rm c}}$ \\
$10^8$ M$_{\odot}$ & kpc & kpc & M$_{\odot}$ pc$^{-2}$ & $10^{8}$ M$_{\odot}$ & kpc  &kpc & 
& $10^8$ M$_{\odot}$ & kpc & kpc & kpc & kpc \\
 \hline
 1.5 & 4 & 0.17 & 8 & 100 & 55.3 & 2.94 & 18.8 & 4 & 4.8 & 1.2 & 2.6 & 0.67 \\
\hline
\end{tabular}
\end{minipage}
\end{table*}

\begin{figure}
\begin{center}
\psfig{figure=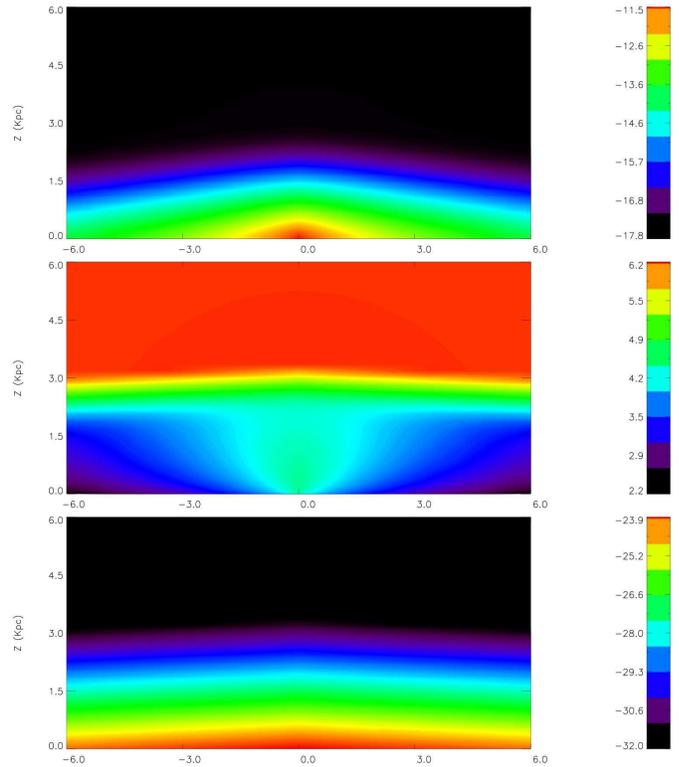,width=0.5\textwidth}
\end{center}
\caption{
Edge-on logarithmic distribution of pressure (top panel), temperature (middle
panel) and density (bottom panel) of the unperturbed ISM in our dwarf galaxy 
model. Distances are given in kpc and the horizontal axis represents either the
x or the y axis crossing the galactic plane. 
Pressure, temperature and density have cgs units, logarithmic scale.}
\label{fig:gal}
\end{figure}

In order to set the gas in a configuration of rotating equilibrium in
the total potential well described above, the pressure at any point is
found by integrating the $z$-component of the hydrostatic equilibrium
equation for any value of the disk radius $R$.  The integration starts
at the outermost values of $z$ (where we can assume the pressure
$P=0$) and proceeds inward to reach the galactic plane at $z=0$.  This
procedure defines the pressure (and hence the temperature) in the
whole space.  Unperturbed initial pressure, temperature and density
distributions for this model are shown in Figure \ref{fig:gal}.

The azimuthal velocity $v_\phi(R,z)$ of the gas is found by balancing
in the $R$-direction the accelerations due to the centrifugal and
gravitational forces as well as to the pressure gradient 
(Figure \ref{fig:star}). 
This method does not guarantee a positive $v_\phi^2$; this must be checked 
{\it a posteriori} \citep[see][]{melioli08}.

\subsection{SFR and SN explosions}
\label{subsec:sfr}

Dwarf irregular galaxies have generally low SFR,
around few $10^{-3}$ M$_{\odot}$ yr$^{-1}$
\citep[e.g.][]{vanzee97, young03, weisz08, lianou13}.  A large
fraction of blue compact dwarf galaxies has instead a higher SFR
ranging between $10^{-3}$ M$_{\odot}$ yr$^{-1}$ and several times 10
M$_{\odot}$ yr$^{-1}$, with a median value of about 0.1 M$_{\odot}$
yr$^{-1}$ \citep[e.g.][]{hopkins02, hunter04, zhao11}; the SF
episodes can last for many hundreds of Myr \citep[][and
references therein]{mcquinn10}.

In order to reach a compromise between the computational cost of the
simulations and the need of a numerical spatial resolution as high as
possible, we consider the SF only in a limited region of the stellar
disk, which we call the ``active region''. This
region extends up to $R=3$ kpc, enclosing 73\% of the stellar mass. 

We calculated three models.  In the first one (MC1) we assume ${\rm
  SFR}=7.3\times 10^{-4}$ M$_\odot$ yr$^{-1}$, and a SF duration
$t_{\rm SF}=250$ Myr. In the second model (MC2) it is ${\rm
  SFR}=7.3\times 10^{-3}$ M$_\odot$ yr$^{-1}$, and $t_{\rm SF}=120$
Myr. The specific SFR (SFR / stellar mass) of these two models is
  $\sim 2 \times 10^{-12}$ yr$^{-1}$ and $\sim 2 \times 10^{-11}$
  yr$^{-1}$, respectively. The former value is somewhat less than
  typical for dwarf irregulars, while the latter lies in the average range
  \citep{weisz11}.

The shorter duration of the stellar formation in MC2 is due to the fact
that we inhibit the SF whenever the mean gas density on the galactic
plane (where the stars are assumed to form) decreases below $10^{-2}$
cm$^{-3}$; this never occurs in model MC1, where the SFR is low and
the gas removal is less conspicuous, but it happens in model MC2 at
about 120 Myr. Of course, once the SNe stop to explode, a back filling
of the ISM takes place increasing the gas density on the plane, and a
second SF burst may arise. However, although we run all our models up
to 250 Myr, we do not consider such a possibility. Finally, we also
calculated a model MC3 in which the star formation is very short
(hereafter indicated as instantaneous), and giving rise to $\sim
2.2\times 10^6$ M$_{\odot}$ of stars. The parameters of the different
models are collected in Table \ref{tab:mod}.

\begin{table}
 \centering
  \label{tab:mod}
 \begin{minipage}{95mm}  
\caption{Models parameters. The rates represent temporal mean estimates during the SN activity.}
  \begin{tabular}{@{}lcccc@{}}
  \hline
      & SFR$^a$& SN rate & SN luminosity & $E^b_{\rm inj}$\\
Model  & M$_{\odot}$ yr$^{-1}$ & yr$^{-1}$ & erg s$^{-1}$ & erg \\
 \hline
 MC1  & $7.3\times 10^{-4}$ & $7.3\times10^{-6}$  & $2\times 10^{38}$
 & $1.75\times 10^{54}$\\
 MC2  & $7.3\times 10^{-3}$ & $7.3\times 10^{-5}$ &$ 2\times 10^{39}$
 & $8.95\times 10^{54}$\\
 MC3  &- & $7.3 \times 10^{-4}$& $2\times 10^{40}$ & $ 2.2\times 10^{55}$\\
\hline
\end{tabular}
\leftline{$^a$ The model MC3 has an instantaneous SF.}
\leftline{$^b$ Total energy released by the SNe during each simulation.}

\end{minipage}
\end{table}

The SN rate is derived from the assumed SFR as: 
${\cal R} = 0.01\times {\rm
  SFR}$ yr$^{-1}$. This is close to the value expected for a Salpeter
IMF with minimum and maximum star mass of 0.1 and 100 M$_\odot$,
respectively. For the model MC3, where the SFR is instantaneous, we
assume that the SNe explode over a timescale of $t_{\rm exp}=30$ Myr,
which is the lifetime of a 8 M$_\odot$ star, the least massive SN
progenitor star.  Thus, the average SN rate is $(0.01\times 2.2\times
10^6)/(3\times 10^7)=7.3\times 10^{-4} $ yr$^{-1}$ for $t\leq 30$ Myr,
and zero otherwise (see Table \ref{tab:mod}).

We note here that the three assumed SN rates are not inconsistent
with the observed SFR-surface density relation.
Given our assumption of 1 SN every 100 M$_{\odot}$ of newly
  formed stars, on the basis of the Kennicutt-Schmidt (KS) relation
  the expected SN rate is \citep{kennicutt98}
\begin{equation}
\label{eq:kenni}
\dot \Sigma_{\rm SN}\simeq 2.5\times 10^{-6}\Sigma_{\rm g1}^{1.4}\;\;\;{\rm
  yr}^{-1}\,{\rm kpc}^{-2},
\end{equation}
\noindent
where $\Sigma_{\rm g1}\equiv \Sigma_{\rm g}/1$ M$_{\odot}$ yr$^{-1}$
is the gas surface density. Taking $\Sigma_{\rm g}\sim 5.5$ M$_{\odot}$
yr$^{-1}$ as the average value within the active area, it turns out that
the SN rate should be $\sim 7.7\times 10^{-4}$, similar to the rate of
model MC3. Even the lower rates of models MC1 and MC2 are compatible
with the observed steepening and dispersion of the KS relation at
low gas surface densities
\citep[e.g.][]{Roychowdhury09,shi11}. This consistency is not
surprising, given that the assumed SFRs lie among the observed range
for dwarf galaxies (see above).

We neglect the supernovae Ia (SNeIa)
contribution from both the energetic and chemical point of view.
Their number is less than half of that of SNeII \citep{mannucci05}, and
the amount of oxygen ejected (the element we are interested in, as a
tracer of the $\alpha$-elements) is
negligible     
\citep[e.g][and references therein]{marcolini07}.

We assume that stars form randomly in time and space, mostly within
stellar associations of different mass $M_{\star,{\rm cl}}$
\citep{watho02}, each generating a total number of SNe $N_{\rm
  SN}=0.01\times M_{\star,{\rm cl}}/M_{\odot}$.  As in
\citet{melioli09}, we assume that the cluster distribution is given by
\citep{higdon05} $f(N_{\rm SN})\propto N_{\rm SN}^{-2}$, where
$f(N_{\rm SN})$ d$N_{\rm SN}$ is the number of stellar clusters with
the number of SN progenitors between $N_{\rm SN}$ and $N_{\rm
  SN}+dN_{\rm SN}$. This distribution is assumed to be valid in the
range $N_{\rm min }<N_{\rm SN}<N_{\rm max}$, with $N_{\rm max}$ = 300,
 while for numerical reasons the value of $N_{\rm min }$ depends
  on the model (see below).

The stars within each cluster are supposed to be coeval, so that the
SNe occur sufficiently close to each other in time and space to
interact mutually, forming a single superbubble. We therefore consider
each association as a single source injecting energy and mass for
$t_{\rm exp}=30$ Myr after the cluster formation at a constant rate of
$L_{\rm w}$ = $E_0N_{\rm SN}/t_{\rm exp}$ and $\dot M = M_{\rm
  ej}N_{\rm SN}/t_{\rm exp}$.  Here $M_{\rm ej}=16$ $M_{\odot}$ and
$E_0=10^{51}$ erg are the mean mass and energy released by a single
explosion, respectively \citep[][and references
therein]{marcolini07}. \footnote{We neglect the stellar wind energy of
  the SN progenitors, as it makes only a minor contribution in the
  total mechanical energy budget, especially for a metal poor stellar
  population relevant for dwarf galaxies
  \citep[see][]{leitherer99}. }

A number of clusters are randomly extracted from the distribution
$f(N_{\rm SN})$, until a total number of supernovae ${\cal R}t_{\rm
  SF}$ is obtained.  In order to place the clusters in space and time
each $i$-th cluster is randomly
associated to a position $P_i$ on the galactic plane ($z$=0) at a
random time $t_i$ in the interval $0<t_i<t_{\rm SF}$. The procedure
generates random points with a spatial frequency proportional to the
stellar density described by equation \ref{eq:rhodisk}. If, however,
the $i$-th cluster happens to be located at distances smaller than 80
pc from a previous association whose SN activity terminated since less
than 60 Myr, its position is recalculated; this is because we assume
that the SN wind of the previous cluster clears the local volume of
its gas, preventing any further SF for a while.

 The above procedure influences the choice of $N_{\rm min}$. In
  principle it can be as low as $N_{\rm min} = 1$, but in model MC1 we
  assume $N_{\rm min} = 5$ because with the disk initial conditions
  assumed here, a superbubble can breakout only when $L > L_{\rm cr}
  \sim 6 \times 10^{36}$ erg $s^{-1}$ \citep{koo_mckee92}, which
  corresponds to $N_{\rm SN} = 5$. However, when the number of OB
  associations is high, after a while the procedure described in the
  previous paragraph hardly allows to place a new OB association, as
  they are too close in space and time. For this reason in MC2 we
  assume $N_{\rm min} = 20$; in so doing we reduce the number of
  sampled clusters. This strategy is physically meaningful; in fact,
  as the smaller associations are more numerous, the bubbles they form
  easily merge producing effects similar to those due to larger OB
  associations. We finally point out that in model MC3 we again assume
  $N_{\rm min}=5$ as all the clusters are located essentially at the
  same time. This variation in $N_{\rm min}$ among the three
  models might have a small impact on the results, since it results
  in a change of the star clusters spatial density which is not
  strictly proportional to the SN rate.

\begin{figure}    
\begin{center}    
\psfig{figure=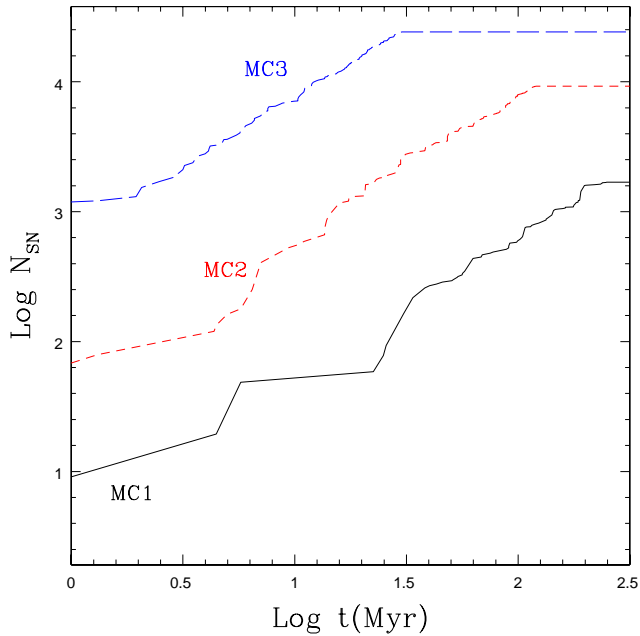,width=0.5\textwidth}    
\end{center}   
\caption{
Cumulative number of supernovae for the three models.
}
\label{fig:sn}
\end{figure}

The cumulative number of SNe
with time for the three models is illustrated in Figure \ref{fig:sn}.

\subsection{Numerics}
\label{subsec:num}

The simulations are calculated using a modified version of the 
adaptive mesh refinement
hydrodynamical code YGUAZU \citep{raga00} that integrates the
gas-dynamic equations using the $flux$ $vector$ $splitting$ algorithm
of \citet{vanleer82}. Radiative losses are taken into account by an
analytic fit to the cooling curve given by \citet{sutherland93} as a
function of the temperature $T$ and metallicity $Z$ for $T \geq 10^4$
K; for $T< 10^4$ K we assume the cooling curve given by
\citet{dalgarno72}. With a highest resolution of 20 pc (see below) the initial 
phase of superbubble expansion is not accurately simulated. 
In particular, the radiative losses would be greatly overestimated due to 
the numerical diffusion which generates large spurious regions of gas at 
intermediate temperatures and density at the boundary between the shocked SN 
ejecta and the shocked ISM. To mitigate this effect we artificially reduce the 
cooling term near the galactic plane. At $z=0$ it is only 5 \% of the real 
value, and gradually increases with the height $z$ to reach the full value 
for $z=300$ pc. We believe that this ``trick'' protects our simulations from 
spurious overcooling while allowing an accurate description of the flow on 
large scale, which is the main purpose of this paper. 
We further investigate the effect of radiative cooling in the Appendix
\ref{sec:appa}.

The 3-D eulerian adaptive hierarchical numerical
grid has five levels of refinements with mesh sizes of 320 pc, 160 pc, 80 pc,
40 pc and 20 pc, respectively. The two coarser grids cover the whole
computational box $-5.2<x<5.2$, $-5.2<y<5.2$, $0<z<10.4$ kpc, while
the finest one extends up to $-3<x<3$ kpc, $-3<y<3$ kpc and $z=160$
pc, the next one up to $-3.84<x<3.84$ kpc, $-3.84<y<3.84$ kpc, 
$0<z<1.28$ kpc, and
the remaining one $-5.2<x<5.2$ kpc, $-5.2<y<5.2$ kpc, $0<z<3.84$ kpc.  In this
study the different grid levels are enforced altogether at any time. 
 A convergence test is presented in Appendix \ref{sec:appb}.

As the grid encompasses only the ``upper half'' ($z>0$) of the galaxy,
outflow boundary conditions are enforced at every grid edge but
the galactic plane ($z=0$), where reflecting boundary conditions are
applied. This latter constraint requires that the SN events must be
located on the midplane; as a consequence we inject into the grid only
half of the mass and energy released by the SNe, as the other half
should be introduced into the ``ghost'' half of the galaxy.
A passive tracer is also added to track the SN ejecta.

The SN energy injection is purely thermal.  Because the numerical grid
is at rest in space while the Galaxy rotates, each OB association
occurring in the active area moves along a circle on the $z = 0$
plane. At each time, we place the energy and mass sources in a single
cell of the finest grid transited by the association at that time.

\section{Results}
\label{sec:result}

In this Section we focus mainly on the dynamics of the ISM and the SN
ejecta. More quantitative discussions on the mass and energy budget and the
chemical enrichment process is postponed to Section \ref{sec:discus}.

In the following we divide the galaxy in two regions: the {\it disk},
defined as the volume within $R<5$ kpc, $z<500$ pc, and the region of
space outside the disk, that for simplicity we
call ``halo''. This separation allows us to characterize the
enrichment of the stellar body, where new stars are formed, and that
of the intergalactic medium (IGM).

For future reference (see Section \ref{sec:discus}) we point out that
the initial amount of unperturbed ISM of the disk within the grid is
$M^{\rm disk}_{\rm g,0}\sim 1.4\times 10^8$ M$_{\odot}$, while the
remaining gas with mass $M^{\rm halo}_{\rm g,0}\sim 10^7$
M$_{\odot}$ is distributed within the halo.

\subsection{MC1}
\label{subsec:mc1}

In the first model (MC1) we assume a continuous mean SFR 
of $7.3\times 10^{-4}$ M$_{\odot}$ yr$^{-1}$ within the active region, 
corresponding to an average SN rate 
of $7.3 \times 10^{-6}$ yr$^{-1}$ within the active
region. This simulation lasts $\sim 250$ Myr, and $\sim 1750$ SNe explode in
this time interval. The SNe are clustered in $69$ different
stellar associations distributed following the power law given in
Section \ref{subsec:sfr}.

\subsubsection{ISM evolution}
\label{subsubsec:ism}

\begin{figure*}    
\begin{center} 
\psfig{figure=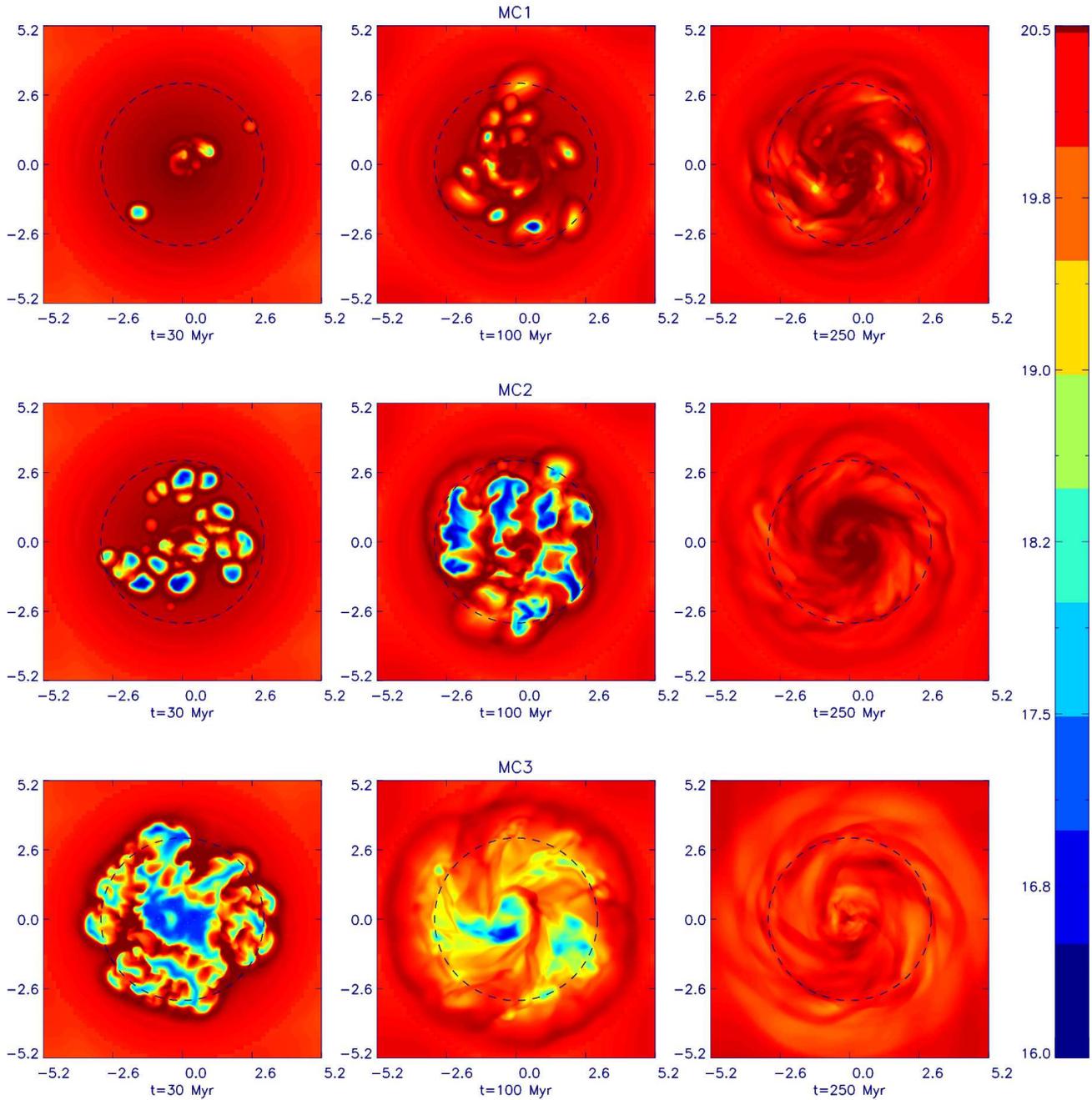,width=1.0\textwidth}    
\end{center}   
\caption{ Distribution of the face-on gas column density (for $0 < z <
  500$ pc). Each
  row refers to a single model. Each column refers to a single
  time. The dashed circle delimitates the active area (see
  text). Distances are in kpc and the (horizontal) x-axis and (vertical) 
  y-axis individuate the galactic plane. 
  Column densities are in log(g$\,$cm$^{-2}$) units. }
\label{fig:zcolumn}
\end{figure*}

The first row of Figure \ref{fig:zcolumn} illustrates the face-on 
ISM column density (calculated for $0<z<500$ pc, that is within the 
disk thickness)
distribution at different times. The holes created
by the SN activity tend to acquire an elongated shape due to
the differential rotation of the galaxy. In general, the largest holes
are created in the outskirts of the active region, where the ISM
is more easily removed owing to its lower density and pressure.  Given
the low SN rate, the filling factor of the bubbles powered by the SN
associations is rather small, and the holes created by the explosions
do not produce a coherent action of gas removal. 

\begin{figure*}    
\begin{center} 
\psfig{figure=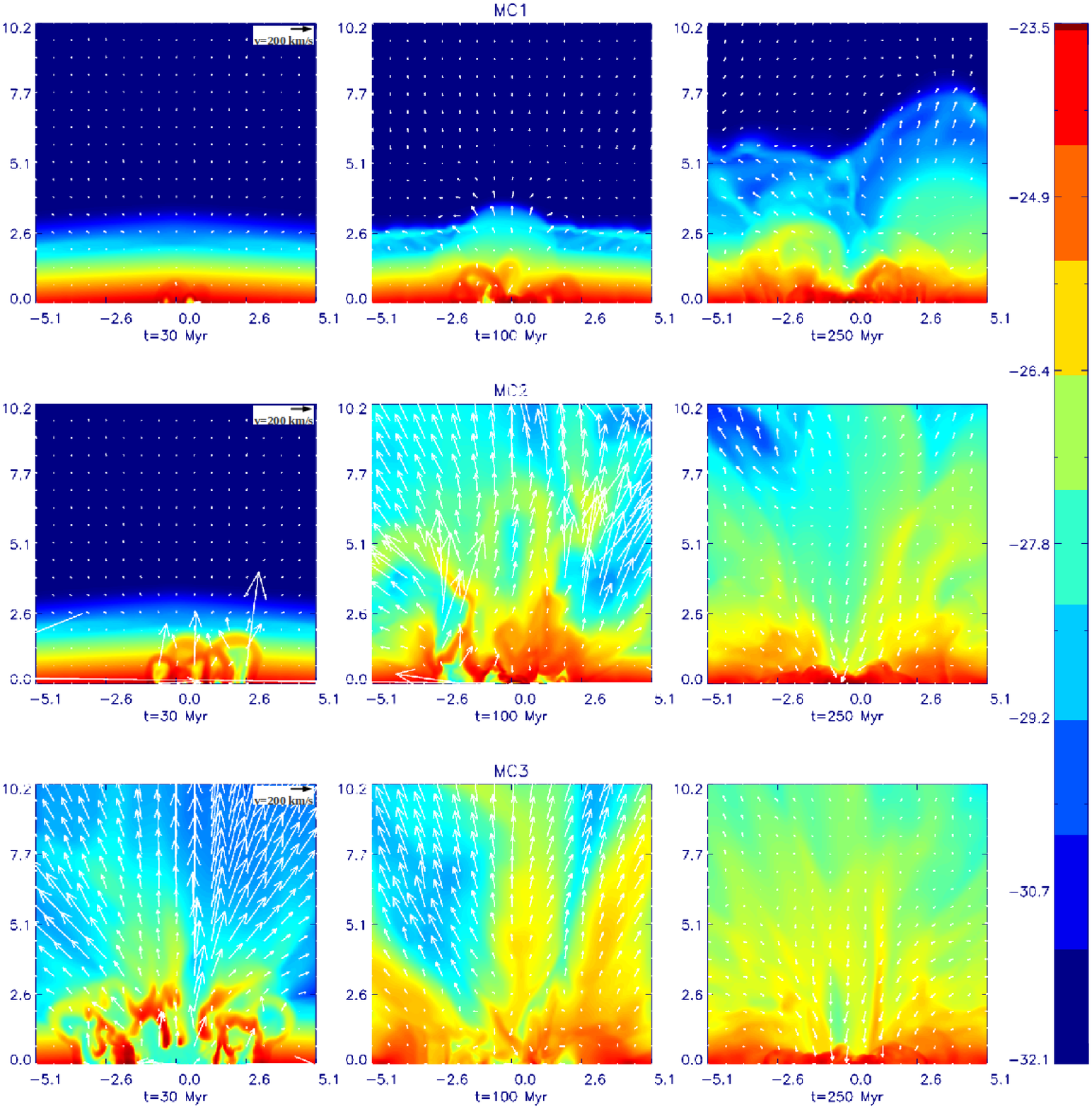,width=1.0\textwidth}    
\end{center}   
\caption{Gas density distribution on the $x=0$ plane (edge-on view). 
Each row refers to a single model. Each column refers to a single time. 
The arrows illustrate the velocity field. Distances are in kpc and the 
(horizontal) x-axis and (vertical) z-axis individuate the plane perpendicular 
to the galactic disk at R=0. Densities are in log(g$\,$cm$^{-3}$) units. }
\label{fig:xcut}
\end{figure*}

Yet, Figure \ref{fig:xcut}
shows that SN clusters can lift gas
well above the disk, although never beyond the grid top boundary.
The filaments and the bubbling structure seen in this
figure are the result of the interaction among shells of the cavities
produced by the single SN clusters. As pointed out above, the low
frequency of the intermittent SN activity prevents an effective
synergy among different bubbles. As a result, while the hot tenuous
ejecta within single bubbles is energized enough to be thrown upward
with velocities that can reach a few hundreds of km s$^{-1}$, the
colder and denser ISM filaments, more resilient to be accelerated,
acquire velocities which hardly rise to 100 km s$^{-1}$.  This is
illustrated in Figure \ref{fig:visto} which shows the mass-weighted
distribution of gas vertical velocity. Anticipating the definition
given in Section \ref{subsec:grad}, in this figure two gas phases are
considered: a hot phase with $T\geq 10^5$ K, and a cold one with
$10^2<T< 10^5$ K. At any time the cold gas moves in the
$z$-direction with a velocity 
$v_z < \sim 85$ km s$^{-1}$, lower than the local escape velocity.
The hot gas velocity distribution, instead, develops a high velocity tail.
The different behaviour of the cold
phase, representative of the star forming ISM, and of the hot phase,
which contains most of the recently produced SN metals, 
is crucial in the chemical
evolution of the galaxy (cf. Section \ref{sec:conclusion}).

\subsubsection{SN ejecta distribution}
\label{subsubsec:ejecta}


\begin{figure*}    
\begin{center} 
\psfig{figure=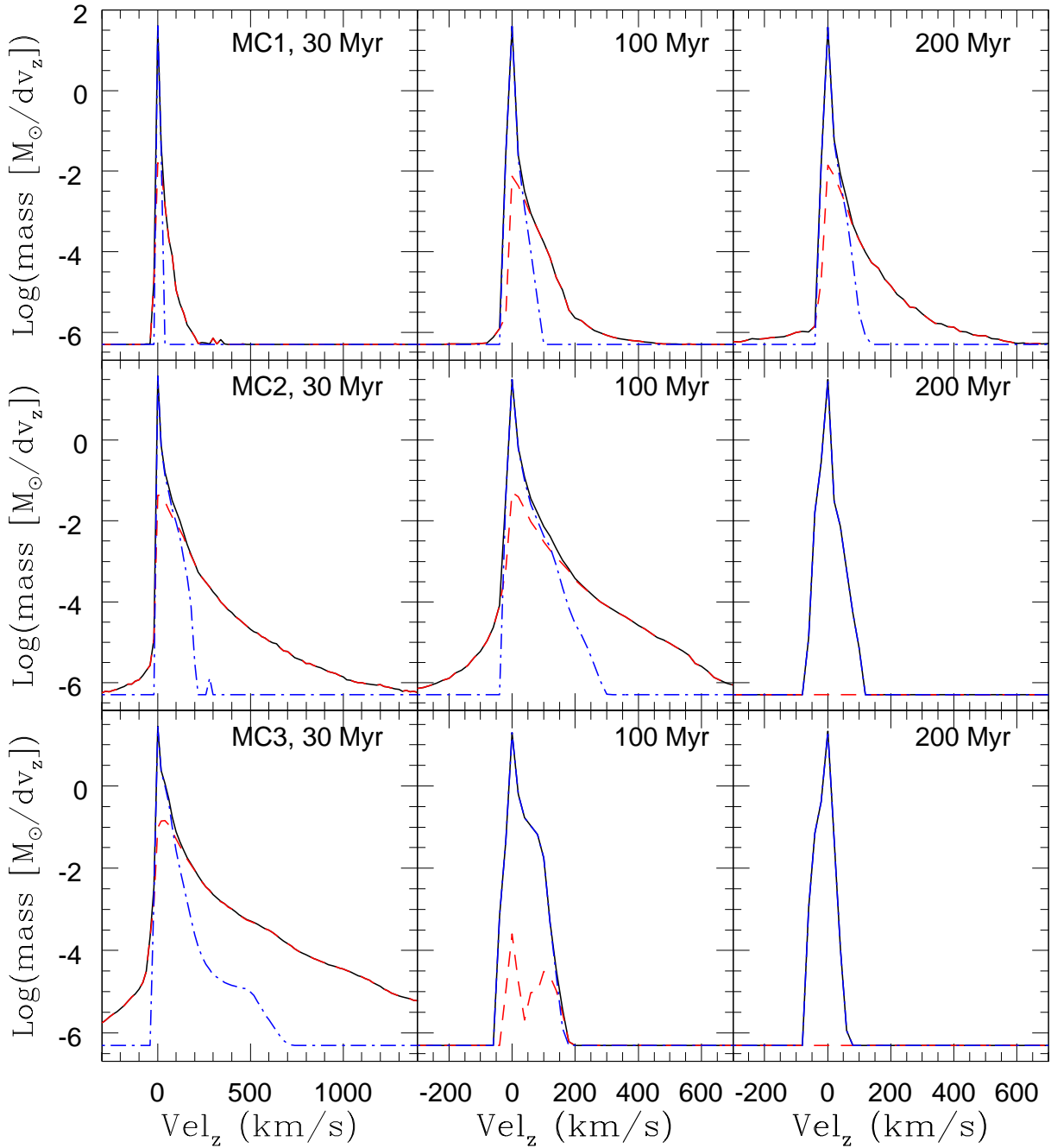,width=1.0\textwidth}    
\end{center}   
\caption{Mass-weighted distribution of gas vertical velocity. The
  dot-dashed line refers to the cold phase ($10^2<T<10^5$ K), the
  dashed line refers to the hot phase ($T>10^5$ K), and the solid
  line refers to the whole gas. Each row refers to a single
  model. Each column refers to a single time. }
\label{fig:visto}
\end{figure*}

\begin{figure*}    
\begin{center} 
\psfig{figure=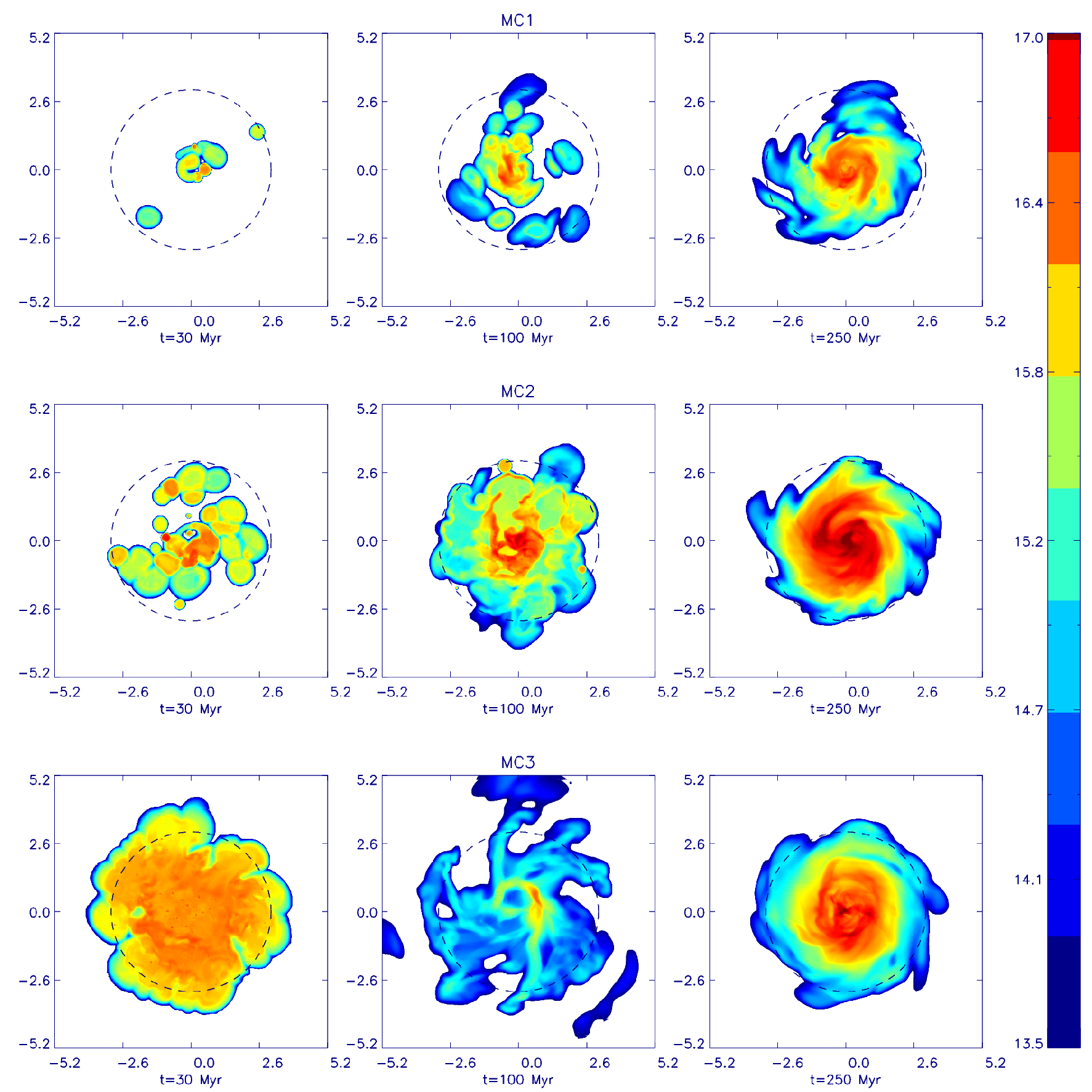,width=1.0\textwidth}    
\end{center}   
\caption{ Column density of the SN ejecta along the $z$-direction in
  the range $0<z\leq 500$ pc. Each
  row refers to a single model. Each column refers to a single
  time. The dashed circle delimitates the active area (see
  text). Distances are in kpc and the (horizontal) x-axis and (vertical) 
  y-axis individuate the galactic plane. 
  Column densities are in log(g$\,$cm$^{-2}$)
  units. }
\label{fig:zcolumn1}
\end{figure*}

\begin{figure*}    
\begin{center} 
\psfig{figure=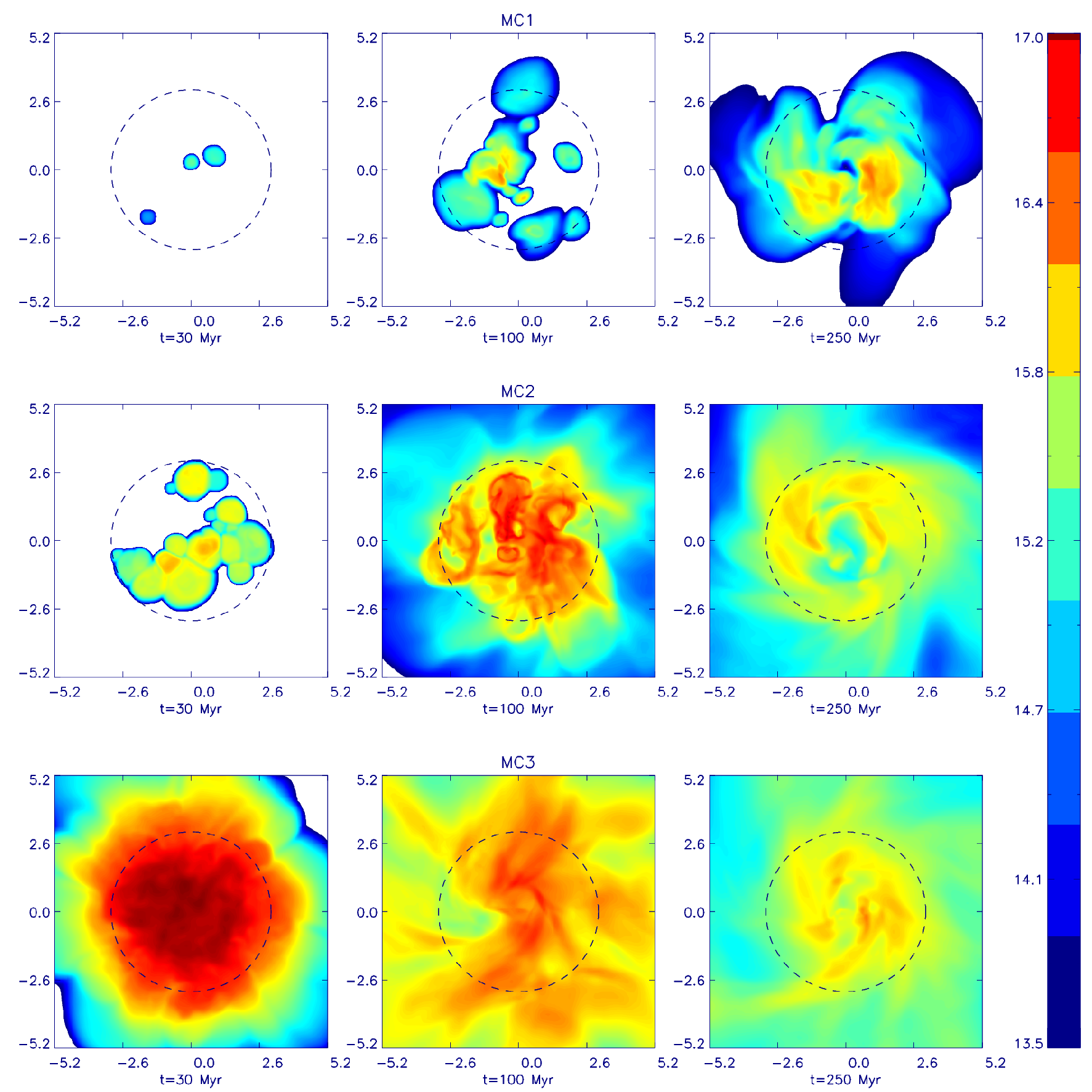,width=1.0\textwidth}    
\end{center}   
\caption{ Column density of the SN ejecta along the $z$-direction for
  $z>500$ pc. Each row refers to a single model. Each column refers to
  a single time. The dashed circle delimitates the active area (see
  text). Distances are in kpc and the (horizontal) x-axis and (vertical) 
  y-axis individuate the galactic plane. Column densities are in
  log(g$\,$cm$^{-2}$) units. }
\label{fig:zcolumn2}
\end{figure*}

Figures \ref{fig:zcolumn1} and \ref{fig:zcolumn2} show the
column density of the SN ejecta along the $z$ direction in the range
$0<z\leq 500$ pc and $z>500$ pc, respectively. These maps nicely illustrate
the circulation of the metals due to the SN feedback. The SN ejecta spatial
distribution within the disk quickly becomes inhomogeneous, with
column density variations of $5-10$ between nearby regions (middle
panel, top row of Figure  \ref{fig:zcolumn1}). With time, rotation and
hydrodynamical dispersal tend to erase azimuthal fluctuations, leaving
a centrally concentrated metal distribution. In the halo, instead,
the SN ejecta has still an irregular distribution at the final time of
the simulation ($t=250$ Myr).  
In Figure 
  \ref{fig:visto} it is shown that only a tiny fraction of the (hot)
  gas is fast enough to leave the system. Evidently, 
the weak SF feedback of model MC1 is only
able to lift some fraction of the SN metals in the halo, where they 
spread over a large volume. The chemical enrichment
process will be discussed in detail in Section \ref{sec:discus}.

\subsection{MC2}
\label{subsec:mc2}

In this model the average SN rate is $7.3\times
10^{-5}$ yr$^{-1}$, an order of magnitude higher than in MC1.  The
occurring 8945 SNe are clustered in $158$ different stellar
associations.

As can be seen in column density map (Figure \ref{fig:zcolumn}),
the galactic ISM on the disk
appears soon punched by holes, which then mostly merge, clearing the
active area of the original ISM. The single bubbles created by
different SN clusters incorporate each other, and their combined
action produces a galactic wind strong enough to lift the ISM to large
heights as can be seen in Figure \ref{fig:xcut}.  An almost steady
  high-velocity tail develops in the velocity distribution
  of the hot component, and even the distribution of the cold ISM
  extends beyond $\geq 250$ km s$^{-1}$ at $t\sim 100$ Myr (see Figure
  \ref{fig:visto}), well above the escape velocity.  The ISM is then
vented in the halo and some of it is able to leave the system, moving
through the upper grid boundary.  
 Contrary to model
MC1, the SN ejecta is also moved to higher values of $z$, as can be
qualitatively understood comparing the column densities in the range
$0<z\leq 500$ pc and $z>500$ pc (middle rows in Figures
\ref{fig:zcolumn1} and \ref{fig:zcolumn2}, respectively). Analogously
to (and more appreciably than in) model MC1, the ejecta in the halo
expands all over the galaxy.

In conclusion, the SNe essentially remove the ISM close to the
equatorial plane. When this happens, at $t\sim 120$ Myr, we stop the
SN explosions, consistently with the assumption that the star
formation process is quenched. We evolved the simulation further, however,
in order to follow the ISM settlement after the SN activity. The holes
generated by such activity shrink in about a sound crossing time. The
disk recovers almost all the gas that was lifted up or pushed
radially, retrieving up to 90\% of the initial mass at $t=250$ Myr,
after 130 Myr since the SN explosions terminated (cf. Section
\ref{sec:discus}).  The dynamical effect of the SF is still observable in the
persistent presence of spiral features in the ISM, as shown in Figure
\ref{fig:zcolumn}. During the quiescent time only a
fraction of the ejecta floating into the halo slowly falls back, while
the rest moves outside the computational grid. The time evolution
of the spatial distribution of the ejecta column density is depicted
in Figures \ref{fig:zcolumn1} and \ref{fig:zcolumn2}. As for model
MC1, significant metal inhomogeneities are present both in the disk
and in the halo, especially
during the SF phase. While in the disk the metal pollution is limited
to the active region, in the halo the SN ejecta reaches larger radial
distances. 

\subsection{MC3}
\label{subsec:mc3}

In this model we consider an instantaneous event of star
formation, in which $\sim$ 3 $\times$ 10$^6$ M$_{\odot}$ of stars form.
The corresponding SN rate is $7.3\times 10^{-4}$ yr$^{-1}$
(c.f. Section \ref{subsec:sfr}).  We have therefore $\sim$
22000 SNe injecting an average luminosity L$_{\rm SN}$ $\sim$
2.2$\times 10^{40}$ erg s$^{-1}$ in a time interval of 30 Myr.

At the beginning, the large number of hot bubbles and the amount of
injected energy are able to disrupt the ISM in the central region of
the galactic plane. Part of the gas is compressed and pushed sideways,
as shown in the first panel of the third row of Figure
\ref{fig:zcolumn}, where the holes appear to be limited by dense
rims. A large fraction of the hot gas acquires vertical
velocities much larger than the escape velocity (cf. Figure
\ref{fig:visto}); this gas leaves the galaxy and crosses
the upward boundary of the computational box (cf. Figure
\ref{fig:xcut}).

At the end of the SN activity, the gas in the dense shells generated
by the superbubbles expansion 
rushes toward the central region on about a sound crossing time ($\sim
200$ Myr). At $t=100$ Myr some holes are still present (see middle
panel of the bottom row in Figure
\ref{fig:zcolumn}).
At later time the galaxy is able to re-grow its gaseous disk, 
as some of the
ISM previously lifted in the halo rains back toward the galactic plane
(see \ref{fig:zcolumn}).
An analogous consideration holds for the
SN ejecta, as it is apparent comparing the last two panels of Figure
\ref{fig:zcolumn1} with those of Figure \ref{fig:zcolumn2}: in fact,
while the ejecta column density in the disk at $t=100$ Myr is
generally lower than at
$t=250$ Myr, the opposite is true for the halo.

As for model MC2, also in this case the SN explosions create spiral
(gas) density waves, still present at $t=250$ Myr.

\section{Discussion}
\label{sec:discus}

\subsection{ISM mass budget}
\label{subsec:ismbudget}

\begin{figure}
\begin{center}
\psfig{figure=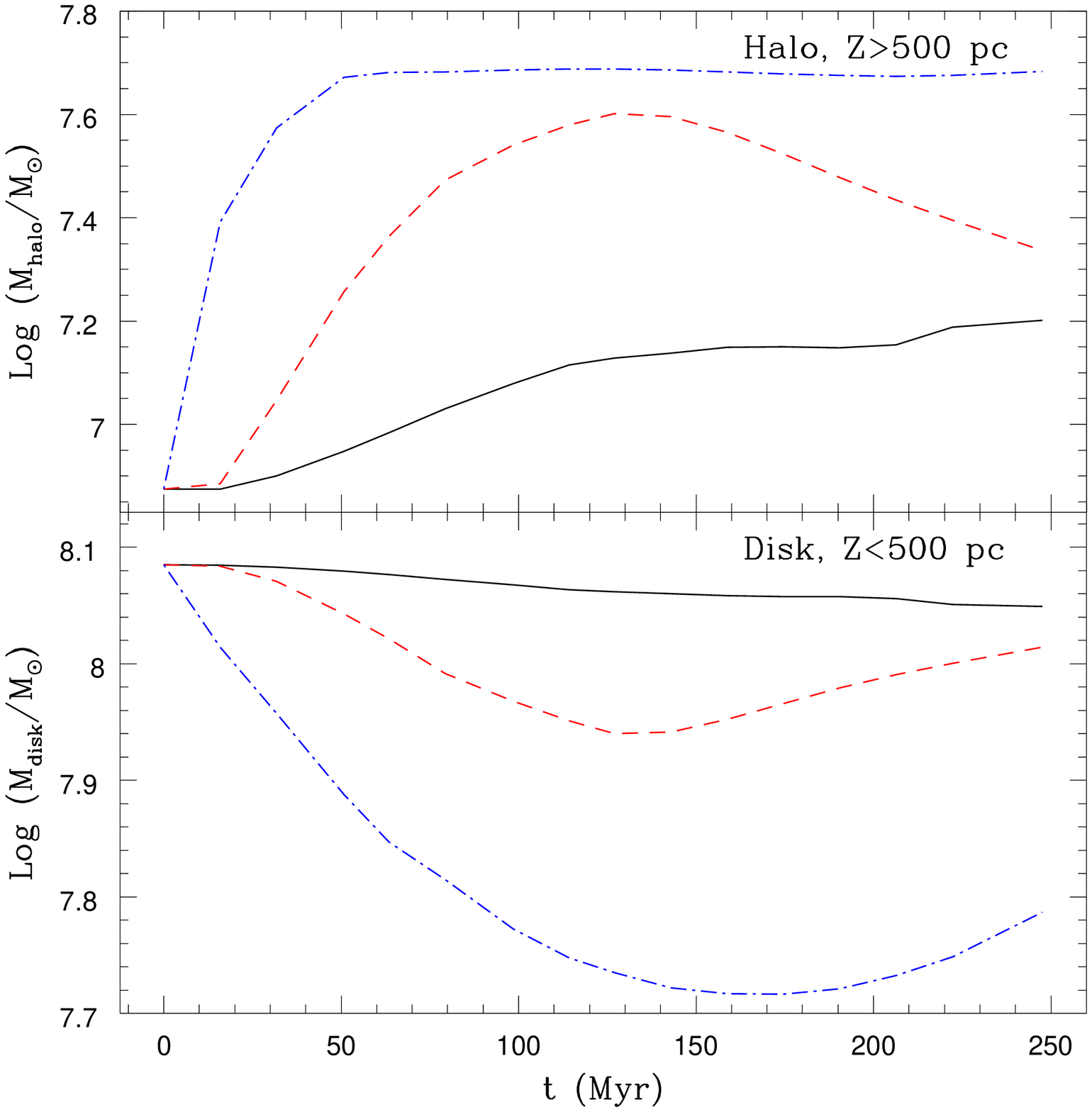,width=0.5\textwidth}
\end{center}
\caption{ Upper panel: time evolution of the mass of gas within the
  halo ($z > 500$ pc) for the models MC1 (solid line), MC2 (dashed
  line) and MC3 (dot-dashed line). Bottom panel: as the upper panel,
  but for the disk ($z < 500$ pc).}
\label{fig:mass}
\end{figure}

Figure \ref{fig:mass} compares the temporal evolution of the ISM
content within the disk and the halo for the three models.

In model MC1 the gas involved in the SN feedback circulation belongs
almost entirely to the active area, as also indicated by Figure
\ref{fig:zcolumn}.  After 250 Myr the disk loses
only $\sim 6$\% of the initial amount of ISM. Comparing the two panels
of Figure \ref{fig:mass}, one sees that this gas is simply lifted into
the halo, and is not lost from the system, as anticipated in
Section \ref{subsec:mc1}.

Because of the higher SN rate, in model MC2 the disk loses a larger
fraction of ISM. The disk gas flows into the halo at an average rate
of $\sim 0.3$ M$_\odot$ yr$^{-1}$, up to the
end of the SN activity, occurring at $t=120$ Myr. At this time the
disk has lost about 30\% of its initial gas content. In the same
period the gas in the halo increases, reaching a maximum at $t=120$
Myr.  After the star formation is halted, the amount of gas in the
disk starts to grow, recovering 85\% of its initial value at $t=250$
Myr.  From the symmetric shape of the dashed curves in the two panels
of Figure \ref{fig:mass} (and from their values) it turns out that, as
for the model MC1, almost all the gas lost by the disk is transferred
into the halo where it remains bounded to the galaxy. When the action of
the SNe ceases, this gas falls back onto the disk at a average rate of
$\sim 0.15$ M$_\odot$ yr$^{-1}$.
At $t=250$ Myr only 2\% of the total initial ISM has been lost.

As expected, the disk of model MC3 suffers the largest and fastest ISM
removal because of the higher SN rate. Interestingly, the disk gas
outflow persists well beyond the SN quenching (occurring at 30
Myr).  In fact, the galactic-scale shock wave, generated by the
cumulative effects of the SN explosions, propagates laterally through
the galactic plane (and also above it) and keeps expanding up to more than
100 Myr, as apparent by the rim present in the second panel of the
third row of Figure \ref{fig:zcolumn} encircling the area
perturbed by the SN activity. As a consequence, the ISM on the plane
is lifted up by this shock, and pushed toward the grid boundaries.
After an initial rapid increase, the amount of gas
within the halo stays approximately constant (cf. Figure \ref{fig:mass}). In fact,
the SN energy release is not spatially homogeneous, and different
funnels form, through which hot and rarefied gas (ISM and SN ejecta)
is channeled upward.  The walls of the funnels, on the other hand,
collide each other, becoming denser and colder because of the increased
radiative cooling due to the compression.  Their rising motion is thus
greatly slowed down and, as a result, they remain bounded to the galaxy,
floating a long time within the halo, as apparent in the last two
panels of the bottom row of Figure \ref{fig:xcut}. As most of the halo gas is
concentrated within the walls, its total amount does not vary at this
stage, as shown in Figure \ref{fig:mass}. At $t\sim 170$ Myr some of
the gas of the halo starts to move back onto the disk. After 250 Myr
the gaseous disk contains 50\% of the initial ISM, while about 40\% of
it is found in the halo. Only $\sim 1.4\times 10^7$ M$_{\odot}$ of gas
have crossed the grid boundaries and left the system. 
This is less than 10\% of the
total amount of gas initially present in the computational box.

Despite the absence of substantial ISM removal, in models MC2 and MC3
the SN feedback significantly alters the distribution of the ISM
between the disk and the halo. An important results is the formation
of massive multiphase ``extra-planar'' gas, located at height $z > $ 500 pc. This
halo gas, generated by the SN action, lasts for several $10^8$ yr and
is analogous to the extra-planar gas generated by galactic fountain in
massive spiral galaxies \citep[][and references therein]{melioli09}.

In conclusion, all the three models retain almost all the initial
ISM. This may be somewhat surprising, as in models MC2 and,
especially, MC3 the total energy delivered by the SNe (cf. Table
\ref{tab:mod}) is comparable with the ISM binding energy, $E_{\rm
  bind} \sim 3.6\times 10^{55}$ erg \footnote{Previous works have
  shown that SNe may substantially fail in ridding the ISM even if
  their energy is much larger than the gas binding energy
  \citep[e.g.][and references therein]{dercole99,marcolini06}.}. 
We discuss this point in Section \ref{subsec:enbudget}.
This result is strengthened by the convergence test presented 
in Appendix \ref{sec:appb}.
There we show that the disk mass content is not strongly sensitive to the
spatial resolution, slowly increasing with the grid refinement.
Therefore the disk gas masses quoted in this Section should be taken
as lower limits.

\subsection{Ejecta mass budget}
\label{subsec:ejbudget}

\begin{figure}
\begin{center}
\psfig{figure=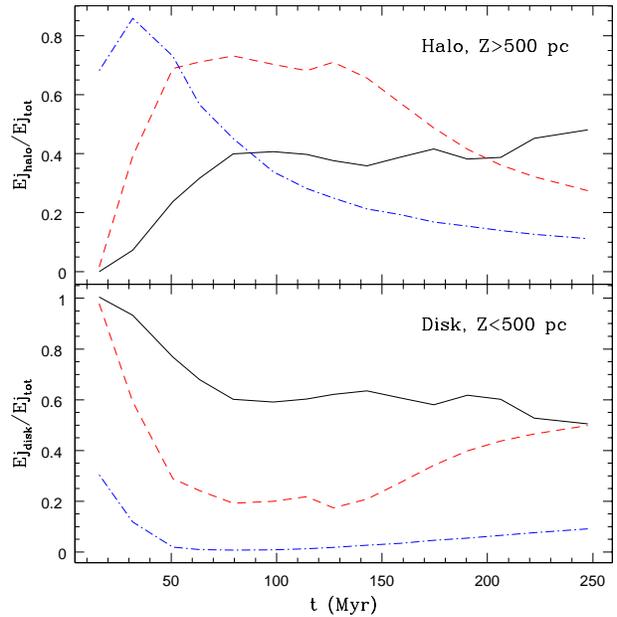,width=0.49\textwidth}
\end{center} 
\caption{ Upper panel: time evolution of the amount of SN ejecta
  within the halo for the models MC1 (solid line), MC2 (dashed line)
  and MC3 (dot-dashed line). Lower panel: as the upper panel, but for
  the disk. At each time the masses are normalized to the mass
  injected by the SNe until that time. In order to go back to the
  absolute values, we report the normalization values at some
  representative times: at times $t=$\{50, 100, 150, 200, 250\} Myr the
  normalization values are $M$=\{4700, 10960, 16640, 22230, 27780\} 
M$_{\odot}$ for MC1;
  at $t$=\{50, 100, 150\} Myr $M$=\{61640, 122520, 141240\} 
M$_{\odot}$ for MC2; after $t=30$
  Myr $M$=383000 M$_{\odot}$ for MC3.
}
\label{fig:ejt}
\end{figure}

As discussed in Section \ref{sec:result}, the fate of
the SN metals is rather different from that of the ISM.  Figure
\ref{fig:ejt} shows the evolution of the SN ejecta mass
(normalized to the total ejecta released at a given time) in the disk
and in the halo.

At early times the freshly delivered ejecta in model MC1 is found
mainly in the disk.  With time, however, the fraction of the ejecta
present into the halo increases at the expense of the fraction
contained within the disk; after 100 Myr an almost steady
configuration is established, in which the amount of new ejecta
lunched upward is approximately compensated by the ejecta in the halo
which tends to move back onto the disk. The relative amounts of
ejecta below and above 500 pc then change rather slowly and, after 250
Myr, are both to $\sim 50$\%, showing that the most of
the metals shed by the SNe are retained by the galaxy.

As long as SNe are active ($t\leq 120$ Myr), the behaviour of the
ejecta in model MC2 is qualitatively similar to that found in model
MC1. Initially, when the amount of the SN ejecta is still quite small,
the ejecta itself is mostly found into the disk; however, after only
$\sim 30$ Myr, its quantity in the halo is about 40\% of the
total. This percentage keeps increasing as long as the SNe explode,
reaching the maximum value of $\sim 70$\%; such fraction is larger
than in model MC1 because of the higher SN rate of model MC2. Once the
SNe stop, the ejecta in the halo starts to fall back onto the disk
where its amount increases, in pace with the behaviour of the ISM
(cf. Figure \ref{fig:mass}).  At the end of the simulation, the
quantity of ejecta within the whole galaxy is $\sim 75$\% of the
total, an evidence that a large fraction of it ($\sim 25$\%) has been
lost by the galaxy.

For model MC3 the SN feedback has a more dramatic impact.
The SN rate is so high that, after
30 Myr (the end of the SN activity) 80\% of the ejecta has been
already transported above the disk.  After this time the ejecta leaves
the halo, with only a tiny fraction of it moving back onto the disk, as
shown in the lower panel of Figure \ref{fig:ejt}. Therefore, in this
model nearly 80\% of the SN metals are lost into the surrounding IGM,
showing that even
dwarf galaxies with moderate SFR are able to enrich the IGM on scales
larger that 10 kpc.

\subsection{Energy budget and SN feedback}
\label{subsec:enbudget}

In order to evaluate the effectiveness of the SN feedback, it is
interesting to compare the total energy released by the stellar
explosions to the binding energy of the ISM, $E_{\rm bind} \sim
3.6\times 10^{55}$ erg.  From Table \ref{tab:mod} we see that $E_{\rm
  inj}/E_{\rm bind}={0.05,\,0.25,\,0.61}$ for models MC1, MC2 and MC3,
respectively. The fraction of lost SN ejecta depends on the injected
energy, its value being ${0.0,\,0.25,\,0.80}$ for the same three models,
respectively. On the contrary, no ISM leaves the galaxy, even in
the most powerful model MC3.These results show that simplistic energy
arguments may be misleading in determining the effect of the SN feedback.

Not all the SN energy goes in fact in venting away the local ISM. As
discussed in Section \ref{subsec:ismbudget}, once the bubbles powered by
the SN associations break out of the galactic plane, channels form
(cf. Figure \ref{fig:xcut}) along which the hot (metal rich) gas is
conveyed into the halo and can leave the galaxy. This is apparent in
Figure \ref{fig:visto} where the high velocity tail present in the
velocity distribution during the SF phase is due to the hot gas
carrying away a fraction of the SN energy in the form of thermal and
kinetic energy; this energy is thus not available for removing the
ISM. We stress that the initial distribution of the ISM plays an
important role. Spherical ISM distributions (presumably the case in
dwarf spheroidal galaxies), make break-outs more difficult to occur,
and most of the SN ejecta could remain trapped within the ISM
\citep{marcolini06}; the gas circulation would then be mainly
regulated by the radiative cooling.

In this case the competition between the rate of the radiative losses
and the rate of the energy injection (by SNe in our models) represents
the key factor regulating the amount of ISM lost by the galaxy. If the
injection rate is low compared to the rate of radiative losses, little
gas is lost by the galaxy, even if the total amount of energy supplied
by the SNe during their activity is higher than the binding energy of
the ISM \footnote{Dwarf elliptical galaxies, with no or little ISM,
  have a dominant early
  burst of star formation, while the gas rich dwarf irregular galaxies
  lack evidence for such an event. \citet{skillman95} suggest that
  this is the cause of the different gas content of the two galaxy
  types, as the SNs explosions associated to the initial SF
  episode are able to remove the
  ISM.}. This is the reason why our models retain most of their ISM
while, for istance, the model ``SMC'' of \citet{hopkins12}, similar in
mass, but with an higher SN rate, suffers a larger mass loss.

\subsection{Chemical enrichment and abundance gradients}
\label{subsec:grad}

One of the main aims of this study is to understand whether the SN
explosions are able to spread the ejected metals over the whole disk,
removing or modifying the intrinsic abundance gradient (given by the
radial distribution of the stars and the ISM --- cf.  Figure
\ref{fig:sdgal}). 

\begin{figure}
\begin{center}
\psfig{figure=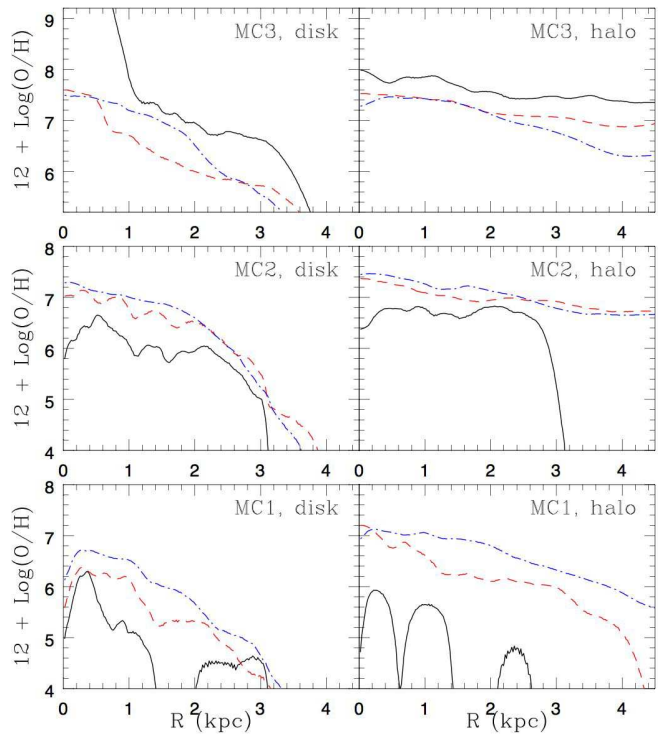,width=0.49\textwidth}
\end{center}
\caption{Radial distribution of the azimuthal average gas metal
  abundance in the disk and in the halo of the three models at three
  different times: $t=30$ Myr (solid line), $t=100$ Myr (dashed line)
  and $t=250$ Myr (dot-dashed line).}
\label{fig:zgrad}
\end{figure}

To this end, at any radius $R$ we have computed the disk gas
metallicity (measured by 12+log(O/H)) defined as the ratio between the
azimuthal average of the oxygen and ISM column densities within
cylindrical shells of radius $R$ and thickness $dR=20$ pc
\footnote{Following \citet{iwamoto99} we assume a mean value of 1.8
  M$_{\odot}$ for the mass of oxygen delivered by a single SN.}.  The
averages are computed between $z_{\rm inf}$ and $z_{\rm sup}$, with
$(z_{\rm inf},z_{\rm sup})=(0,0.5)$ kpc and $(z_{\rm inf},z_{\rm
  sup})=(0.5,5)$ kpc for the computation within the disk and the halo,
respectively.  Figure \ref{fig:zgrad} illustrates the radial profile
of metallicity in the disk and in the halo for the three models at
three different times.  We stress that we consider here only the
metallicity contribution of the simulated SF episode, neglecting the
initial metal content of the ISM. In this way, we emphasize the role of
recent SN feedback in shaping the disk abundance gradient.

The ``holes'' shown at early times in the radial distribution of the
metallicity (e.g. the one present in model MC1 for
$1.4<R<2$ kpc, see Figure \ref{fig:zgrad}) are due to the absence of
recent SN remnants at that radius (cf. the first panel of Figure
\ref{fig:zcolumn2}). At later times, when a nearly steady gas
circulation is set in (see Section \ref{subsec:ejbudget}), the
metallicity distribution assumes a smoother profile. We note that the
solid line (corresponding to $t=30$ Myr) is the lowest one for models
MC1 and MC2, but the highest for MC3. In fact, the SNe explode during
an extended period in the former two models, gradually increasing the gas
metallicity with time. Conversely, in model MC3 all the metals are
released very soon, and are then progressively lost by the galaxy;
this explains the reduction of the dashed line level ($t=100$
Myr). However, at later time the metallicity within the active area of
the disk rises again as the metal-rich gas in the halo cools and fall
back onto the disk (cf. Figs.  \ref{fig:mass} and \ref{fig:ejt}).  The
main result emerging from Figure \ref{fig:zgrad} is that a radial
gradient of metallicity within the disk is substantially always
present in every model, preserving, at least in part, the gradient of
the stellar (and SN) distribution (cf. Figure \ref{fig:sdgal}); on the
contrary, only shallow gradients are present in the extra-planar gas
located in the halo of all the
models. Evidently, given its higher sound speed, the hot, low density
gas floating above the disk is mixed by turbulence more
effectively than the gas in the disk.

\begin{figure}
\begin{center}
\psfig{figure=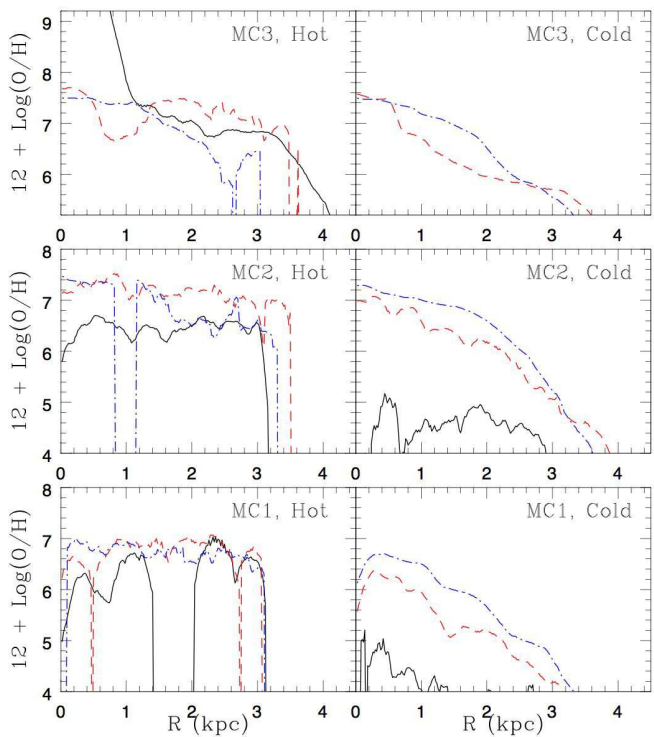,width=0.49\textwidth}
\end{center}
\caption{Radial distribution of the azimuthal average gas abundance of
  the metals in the disk of the three models at three different times:
  $t=30$ Myr (solid line), $t=100$ Myr (dashed line) and
  $t=250$ Myr (dot-dashed line). Left panels refer to the hot
  phase ($T>10^5$ K), and right panels refer to the cold phase
  ($10^2<T<10^5$ K).}
\label{fig:zgradph}
\end{figure}

The abundances shown in Figure \ref{fig:zgrad} are mass-averaged values
within the whole galactic region. Actually, two gas phases coexist
within this region: an hot phase with $T\geq 10^5$ K, and a cold one
with $10^2<T< 10^5$ K \footnote{The chosen temperature threshold of
  $10^5$ K separating the two phases is somewhat arbitrary. However,
  the results shown here do not depend strongly on such a choice; in
  fact, as the maximum of the cooling curve occurs around $T\sim 10^5$
  K, the gas with temperatures up to this value cools rather quickly,
  becoming incorporated in the ``cold'' phase independently of the
  exact value of the chosen threshold temperature
  \citep[c.f.][]{recchi13}.}.  As we are interested in the
possible role of the SN ``fountains'' in flattening the radial
distribution on the disk of the stellar chemical abundance, we focus
now on the cold ISM phase in the disk, out of which new stars will
form. Figure \ref{fig:zgradph} illustrates the disk radial profile of
the abundance of the cold and hot ISM phases. After a comparison with
Figure \ref{fig:zgrad}, it turns out that at early times ($t=30$ Myr),
especially for models MC2 and MC3, the metals reside essentially in
the hot phase; this must be expected, as the SN ejecta is found within
the hot interiors of superbubbles and chimneys. At later times,
however, the ejecta cools down and accumulates into the cold phase
which therefore becomes predominant in shaping the total metal
distribution. This distribution preserves a substantial gradient, and
we thus conclude that the gas circulation due to the SN fountains is
ineffective in redistributing the metals and in homogenizing the
abundance of the cold ISM and of the next generations of stars.  The
uniform abundance distribution in the disk of the hot phase
established at later times (left panels in Figure \ref{fig:zgradph})
would not play any role even if this component would cool at a
subsequent stage, because the amount of the hot ejecta at this time is
negligible.

  As discussed in Appendix \ref{sec:appa}, the numerical diffusion
  at the contact surfaces separating the hot/cold phase determines an
  excess of radiative losses. Our tests show that, even under extreme
  assumptions, the amount of cold gas within the disk does not vary by
  more than a factor of 2, and its dynamics is not greatly
  affected. The amount of hot gas and ejecta is instead more sensitive to the
  cooling prescription adopted.
  This, in turn, implies that the radial abundance profiles in the hot phase are
  affected by radiative cooling, differing by up to a factor of $\sim
  10$ between our standard runs and analogous models without
  radiative losses. Of course, the latter models are rather extreme and
  have been calculated only to show the maximum range of variation.
  However, given the small amount of hot phase ejecta involved (compared to 
  the cold gas), the chemical gradients in the cold disk (which are those 
  influential for the chemical characteristics of the successive generations
  of stars) persist even in this case, showing that their presence is a robust 
  result of our models.

  In Appendix \ref{sec:appb} we tested our results against the
  numerical resolution. We find that the 
  amount of gas within the disk increases with the grid refinement,
  but not by a large amount. 
  Our test shows that the disk ISM mass
  is only $\simlt 15$\% larger when the
  zone size is decreased by a factor of two. The ejecta mass in the
  disk is also quite stable:
  improving the resolution to $10$ pc its mass in the disk increases
  by less than 25\%.  As a result, the chemical gradients are fairly
  insensitive on the numerical resolution adopted.


\subsection{Flattening of chemical gradients}
\label{subsec:flat}

As no relevant gradient is observed in dwarf galaxies, some mechanisms
must be at work in various combinations to distribute
metals. Turbulence has been accepted as the major process shaping
interstellar medium \citep[cf. the reviews by][and references
therein]{elmegreen04,scalo04,burkert06}. However, turbulence has been
shown to dissipate quickly if the external driving is stopped,
resulting in the need to continuously drive turbulence either locally
via stellar feedback or globally via, e.g., clumpy infall of cold gas,
spiral density waves, tidal torques, differential rotation,
magnetorotational instabilities. SN feedback is likely the most important
contributor to ISM turbulence
\citep[e.g.][]{maclow04}. However, our models indicate that this mechanism
is not effective in mixing metals on galactic scales.

Consider, for instance, our model MC1: the SN turbulence do not erase
the chemical gradient, originated by the radial distribution of the
SNe, which is continuously restored by new explosions
\citep[cf.][]{martin95}. However, in case of an intermittent star
formation history, other drivers among those listed above could play a
role during the inactive phases, as suggested by the flattening of the
metallicity gradients in the outer parts of relatively quiescent
spiral galaxies \citep{martin95,bresolin09}.

Taking the sound speed $\sim 10$ km s$^{-1}$ as an upper limit to the
velocity at which mixing occurs, it turns out that $\sim 100$ Myr is a
lower limit to the timescale over which chemical gradients are erased
over 1 kpc range. Therefore, if the star formation proceeds by bursts
separated by quiescent periods of a few hundreds of Myr, as suggested
by recent works on H II galaxies \citep[][and references
therein]{lagos12}, turbulence may well be the cause of flat chemical
gradients.

A further mechanism to flatten chemical gradients has been
proposed recently by \citet{schroyen11}. These authors point out that 
rotation is a key factor to keep the ISM chemically homogeneous in
dwarf galaxies. The resulting ``centrifugal barrier'', in fact,
prevents the gas from collapsing to the dense central region. This, in
turn, results in a more spatially uniform SFR and, consequently, a flatter
stellar mass and abundance profile. This agrees with the observations
by \citet{koleva09}, who found that dwarf galaxies without stellar
population gradients are also the fastest rotating ones.

For more continuous star formation regimes, the metal mixing, as
discussed above, does not happen fast enough to wipe out the
metallicity gradients. A similar conclusion is reached by
\citet{werk11} about the absence of metallicity gradients in the
outskirts of typical star-forming spiral galaxies. In these cases, a
clear explanation for the flatness of the chemical distribution is
still missing. Given the fast mixing speeds required, the metal
transport may be occurring predominantly in a hot gas component, as
proposed by \citet{Tassis08}. Our models provide only limited support for
this scenario. The abundance gradient in the hot gas in the halo is
essentially flat, but to be truly effective this mechanism
requires that the majority of the ejecta is in the hot halo phase,
where its abundance homogenizes and rains back into the disk.
Instead, in our models about half of the SN metals always
resides in the cold gas within the disk, exhibiting a negative
gradient. 
Unfortunately, due to numerical
diffusion, the amount of the ejecta which remains confined within the
disk during the burst is likely overestimated in our
models. Higher resolution simulations are needed to evaluate
more accurately the amount of metals that can be transported into
the hot phase, and then quickly cooled to reproduce the observed flat
chemical distribution.

Finally, we point out that, in addition to the physical processes
previously outlined, there is a further mechanism usually neglected,
but likely relevant in this context. \citet{marcolini04} showed that
the amount of the SN ejecta retained by the galaxy is quite sensitive
to the ram pressure due to the motion of the galaxy through the
intracluster or intragroup medium. In fact, the ram pressure can push the 
ejecta ``floating'' in the halo back toward the disk, increasing the 
fraction of the metals trapped into the disk by a factor depending on
the orientation of the galaxy with respect to the
direction of motion. The disk enrichment process is thus also affected by
the environment in a complex way, not investigated here.


\section{Summary and Conclusions}
\label{sec:conclusion}

As discussed in Section \ref{subsec:dwarf}, we considered a dwarf galaxy
with the initial ISM distribution in rotational equilibrium within a
gravitational well due to both the stellar disk and the spherical
dark matter halo. The aim of the paper was to study the large scale
gas flow induced by SN feedback during the ``quiescent'' (non
star-bursting) stage representing the largest phase of the life of dwarf
galaxies \citep{lee09}, and three regimes of star formation were considered. In
particular, we focused on the circulation and distribution of the
metals delivered by the SNe, in an attempt to reproduce substantially
flat metallicity gradients as those observed.

In agreement with previous theoretical studies \citep[e.g.][]{maclow99,
  dercole99, melioli13}, the absence of substantial ISM loss is
verified in our models, even when the SN energy injected
is comparable to the ISM binding energy. This is because the radiative
cooling is effective in neutralizing the SN heating and because the
flattened distribution of the ISM does not allow an efficient coupling
between the SN energy and the gas.

Despite the absence of a sizeable gas removal from the galaxy, the SN
feedback has nevertheless a profound impact in the galactic
evolution. In fact, even a relatively weak SFR, as the one of model
MC2, is able to temporarily lower the ISM density and stop the SF process.
At the end of the SN activity, the expelled ISM begins to slowly fall
back to the disk, re-building a massive, cold ISM in $\approx 10^8$
yr, possibly giving rise to a new SF episode, and leading
to an intermittent SF history.

For the weakest SFR considered here, model MC1, the SN feedback is
instead unable to drastically affect the ISM. In this case the SN
associations generate holes in the gaseous disk but do not evacuate
completely the active region, and the star formation can
continue for long time at a similar rate.

For all the models discussed here the flow triggered by the SN heating
generates a long-lived extra-planar gas, especially conspicuous for models MC2 and
MC3, as those actually observed in dwarf galaxies.

The circulation of the SN ejecta is rather different from that of the
ISM. When the bubbles break out, some or most of their hot interior is
expelled with velocities larger than the escape velocity. As a result,
after 250 Myr the SN ejecta present in the galaxy (both in the disk
and the halo) is anticorrelated to the SN rate, being 100\%, 75\% and
20\% of the total amount for models MC1, MC2 and MC3, respectively.
The trapped ejecta is nearly equally accommodated in the disk and in
the halo for all the models.

Despite the reduced size and the shallow potential of the dwarf
galaxies, the SN ejecta does not spread all over the disk, and does not
give rise to the observed nearly flat metal distribution.  Instead,
the models presented here show the existence of persistent chemical radial
gradients as in Milky Way-sized galaxies,
where galactic fountains do not lead to efficient radial metal mixing
\citep{melioli08,melioli09}. In Section \ref{subsec:flat}
we discussed several mechanisms which could be at work in flattening the
chemical gradients, and their compatibility with our models.

\section*{Acknowledgments}
We are grateful to the unknown referee whose suggestions and
criticisms substantially improved the paper.  CM acknowledges financial
support from grants from the Brazilian Agencies FAPESP.  FB is
supported in part by the Prin MIUR grant 2010LY5N2T``The Chemical and
Dynamical Evolution of the Milky Way and Local Group Galaxies''.

\bibliographystyle{mn2e} \bibliography{dw_ref}

\appendix
\section{Radiative Cooling Effectiveness}
\label{sec:appa}

In Section 4.3 we highlighted the importance of the
radiative cooling in regulating the amount of the cold ISM, out of
which new generations of stars can form. Unfortunately, an unavoidable
limit of the numerical simulations is given by the inaccurate
reproduction of the contact discontinuities separating the hot and
cold phases such as those between the hot bubble interiors and the
cold shells. In fact, the numerical diffusion tends to smear such
discontinuities creating regions of intermediate temperatures and
densities; as a consequence, the radiative cooling and the mixing of
the hot metals with the surrounding cold ISM are artificially
increased \footnote{Actually, a physical spread of the contact
  discontinuities is expected owing to the presence of thermal
  conduction and mixing layers, but, in general, it remains too thin
  to be properly resolved by large scale numerical simulations
  \citep[c.f.][]{recchi01,marcolini04,parkin10}}.

Underestimates as well as overestimates of the radiative cooling may
also occur when a cooling curve calculated in a
regime of ionization equilibrium is adopted (as we actually do). A
hot plasma may cool at lower rates \citep{sutherland93} or at higher
rates \citep[e.g.][]{borkowski90} relative to the condition of
ionization equilibrium, depending on its thermal history.

\begin{figure*}    
\begin{center} 
\psfig{figure=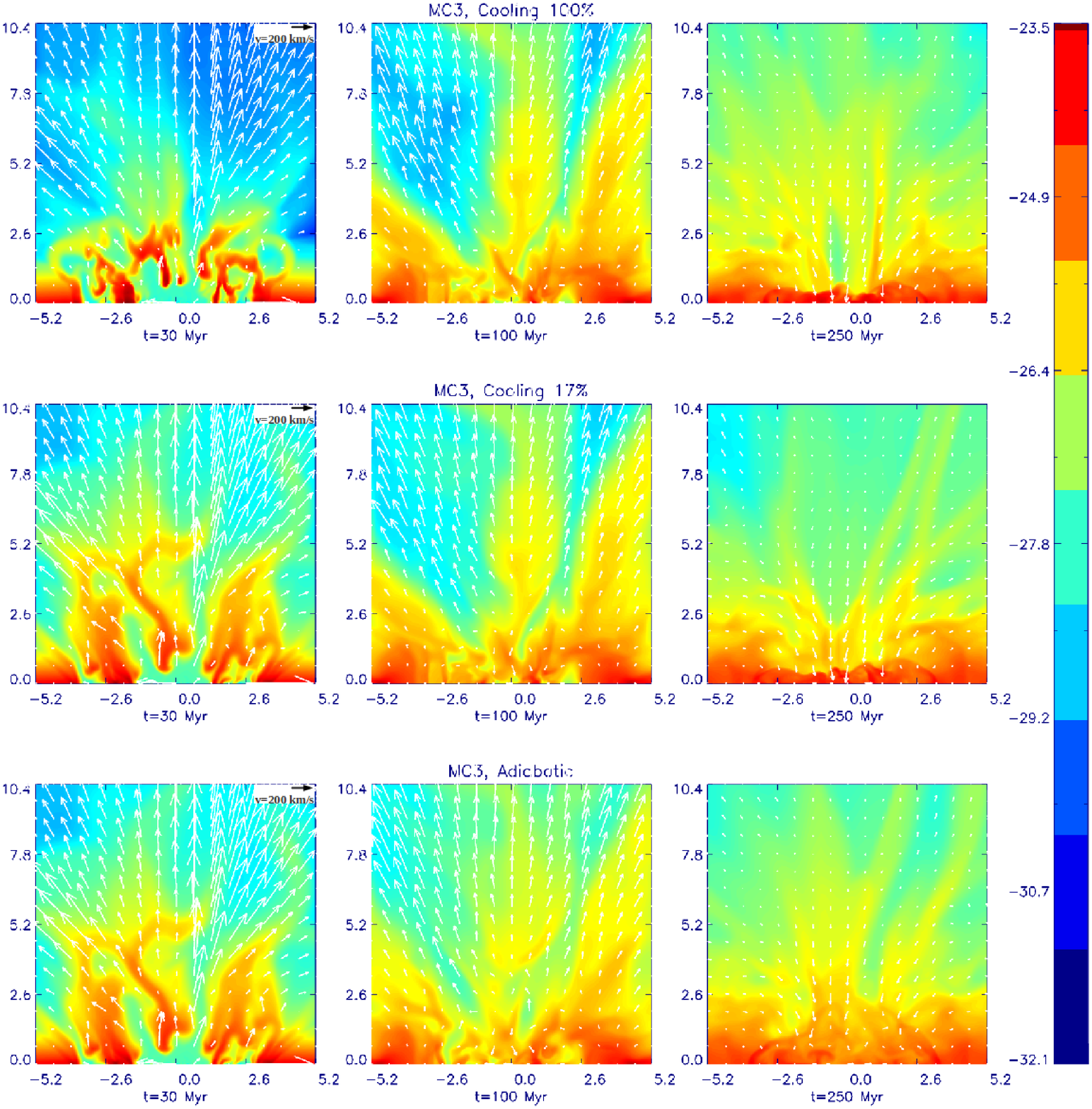,width=1.0\textwidth}    
\end{center}   
\caption{ Gas density distribution on the $x=0$ plane for model
  MC3. From top to bottom, the rows refer to a radiative cooling
  efficiency $\eta=1$, $\eta=0.17$ and $\eta=0$, respectively. Each
  column refers to a single time. The dashed circle delimitates the
  active area (see text). Distances are in kpc and the 
  (horizontal) x-axis and (vertical) z-axis individuate the plane perpendicular 
  to the galactic disk at R=0. Densities are in log(g$\,$cm$^{-3}$) units. }
\label{fig:xcutrad}
\end{figure*}

In order to ascertain the sensitivity of the results to the radiative
effectiveness, we run two additional models of MC3 in which the cooling
term is multiplied by the factor $\eta=0.17$ and $\eta=0$ (adiabatic
case), respectively.  Figure \ref{fig:xcutrad} shows the gas
distribution on the $x=0$ plane of the two models, as well as of the
standard model with $\eta=1$. It is apparent that the models with
$\eta=1$ and $\eta=0.17$ are quite similar. We conclude that, at least
for a general description of the gas circulation, the exact shape of
the cooling curve is not crucial and the rate of radiative losses is
more dependent, on (the square of) the gas density.

\begin{figure}
\begin{center}
\psfig{figure=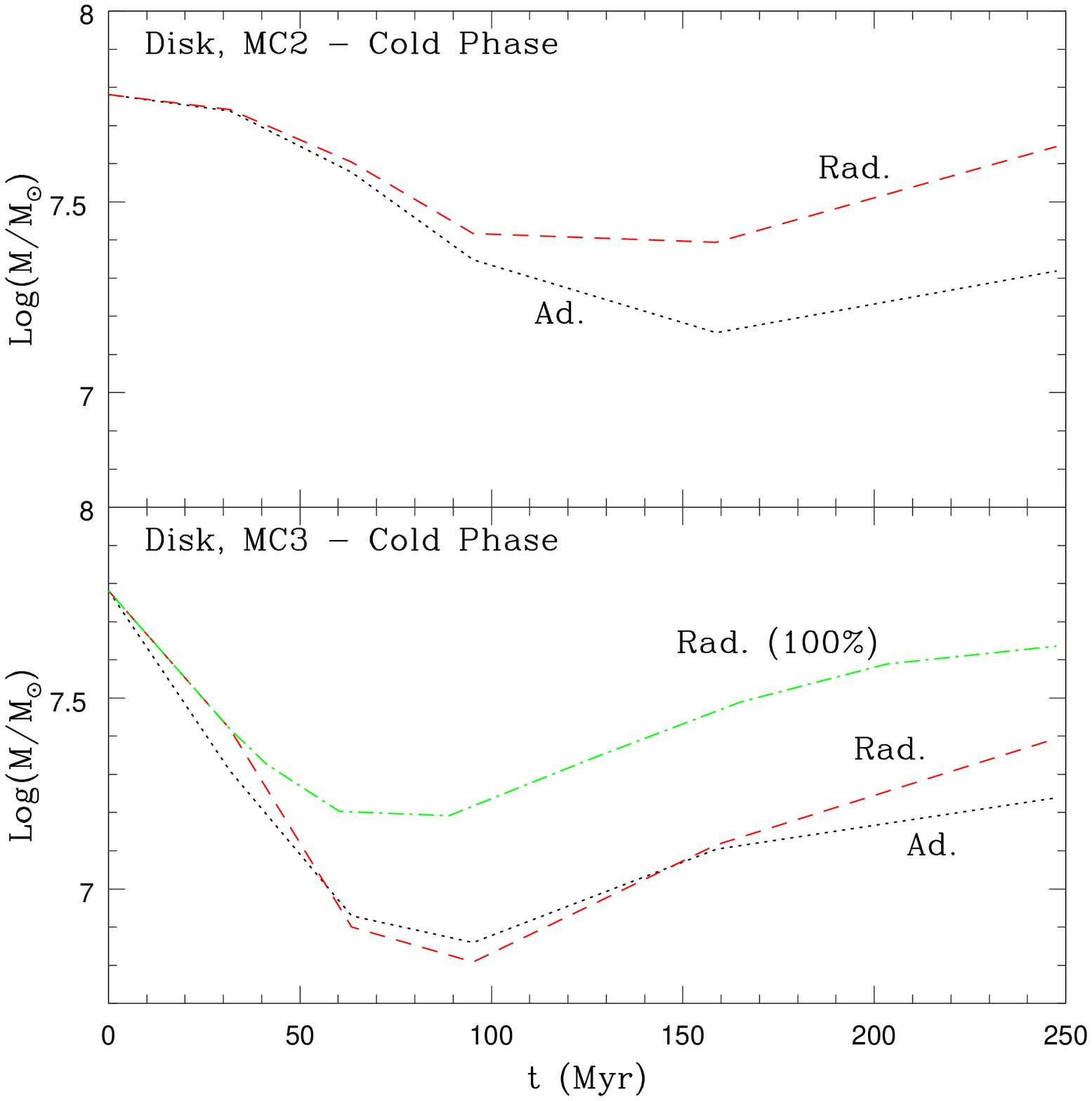,width=0.49\textwidth}
\end{center}
\caption{Time evolution of the amount of the cold ($T \leq 10^5$ K)
  ISM phase within the disk for model MC2 (upper panel) and MC3 (lower
  panel). The dashed and dotted lines refer to the cases $\eta=1$ and
  $\eta=0$, respectively. The dot-dashed line in the lower panel,
  labelled as ``100\%'',
  represents the mass evolution once our assumption on the dependence
  of the cooling rate on $z$ (cf. Section \ref{subsec:num}) is
  dropped. Masses are given in solar masses.}
\label{fig:cold}
\end{figure}

A more quantitative evaluation of the effect of the radiative
effectiveness is shown in Figure \ref{fig:cold}. In the two panels the
amounts of cold gas ($T\leq 10^5$ K) in the disk region where the SN
explosions occur (that is, in the region defined by $R<3$ kpc; $z<
500$ pc) for models MC2 and MC3 are
compared for the cases $\eta=1$ and $\eta=0$.

One can see that, in model MC3, the cold gas evolution in the adiabatic case
follows the same ``trajectory'' of the full radiative case, and starts
to deviate from it only at $t \sim 160$ Myr; after 250 Myr the amount of
cold gas in the adiabatic model is 40\% lower than in the fully
radiative case. The model MC2 follows the same qualitative behaviour,
but the adiabatic and the radiative models start to differ
after $\sim 90$ Myr; at $t=250$ Myr the cold gas in the adiabatic
model is nearly half than in the radiative model. This larger
discrepancy is due to the fact that in model MC3 the higher SN
luminosity heats the gas to higher temperature, which results in a longer 
(average) cooling time. Moreover, many superbubbles coalesce,
reducing the global extension of the contact surface between hot and
cold ISM (where most of the radiative losses occurs). In model MC2,
where the SN luminosity is lower and prolonged, the heated gas
acquires a lower temperature and, consequently, a shorter cooling
time.  In addition, the weaker SN feedback allows the survival of both
cold and hot gas in the active region; this leads to an
increase of the contact surfaces. Both effects exacerbate the spurious
overcooling. 

As a further test we run the model MC3 with $\eta=1$ and dropping our
assumption on the dependence of the cooling rate on $z$
(cf. Section \ref{subsec:num}). The result is illustrated by the dot-dashed line in the 
lower panel of Figure \ref{fig:cold}. In this case the amount of cold mass is higher
relative to the ``standard'' model by a factor within 1.5 -- 2.5, which is an
overestimate of the realistic value (due to numerical overcooling).

\begin{figure}
\begin{center}
\psfig{figure=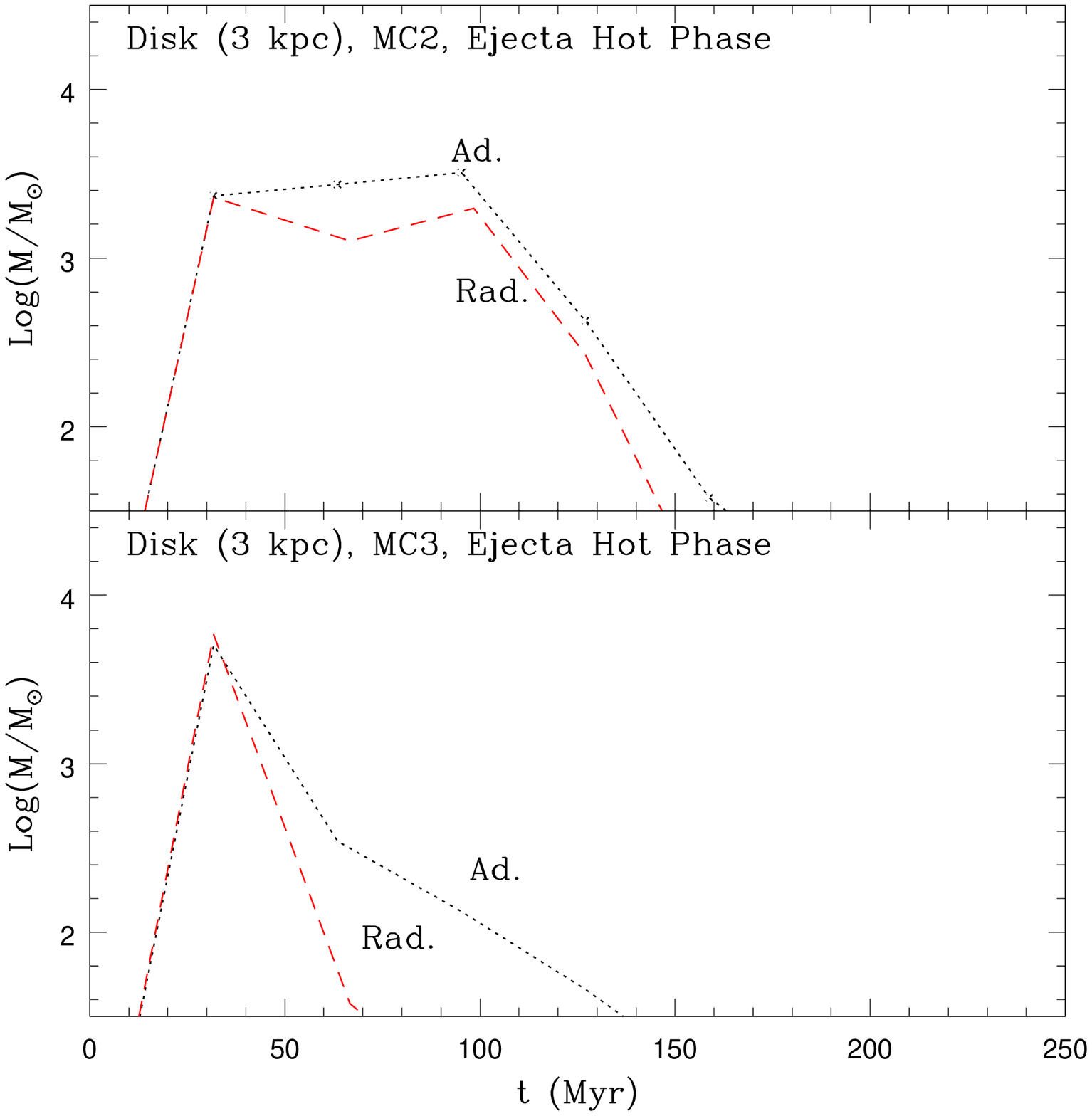,width=0.49\textwidth}
\end{center}
\caption{Time evolution of the amount of the hot ($T> 10^5$ K) ejecta
  phase within the active area of the disk ($R
\leq 3$ kpc) for model MC2 (upper panel) and MC3 (lower
  panel). The dashed and dotted lines refer to the cases $\eta=1$ and
  $\eta=0$, respectively. Masses are given in solar masses.}
\label{fig:hot}
\end{figure}

The above discussion shows that numerical uncertainties in the radiative losses
likely do not strongly affect the amount and the dynamics of the cold gas in
the disk, unless a prolonged, low intensity SN activity is going
on. Sadly, the situation is more complicated for the processes of
metal dispersal and mixing.  Figure \ref{fig:hot} shows the evolution
of the mass of the SN ejecta in the hot ($T>10^5$ K) phase located in
the disk region (and bounded by $R<3$ kpc), for the standard models
MC2 and MC3 and the corresponding adiabatic analogues. For MC2 the
difference in $M_{\rm ej}^{\rm hot}$ is moderate, less than a factor
of 2 for most of the time. This relatively small difference is partly
due to our choice to artificially reduce the radiative losses near the
galactic plane (cfr. Section 2.3). However, the continuous energy
injection by SNe, characteristic of this model, also helps to mitigate
the overcooling problem described above.

In fact, for model MC3, the difference between the standard and the
adiabatic run is more dramatic. At $t \sim 60$ Myr the discrepancy in $M_{\rm
  ej}^{\rm hot}$ amounts to about a factor of 10.

This comparison between extreme models (radiative and adiabatic) gives
a measure of the uncertainties due to the inaccurate numerical description of
the surfaces separating hot and cold gas. While this error seems
acceptable for models MC1 (not shown here) and MC2, we are not in the position
to make strong statement for model MC3, for $t \geq 60$ Myr. Higher
resolution simulations (in preparation) will shed light on this issue.

\section{Convergence Test}
\label{sec:appb}
Here we describe how the grid resolution influence our results, in
particular the amount of gas mass in the disk which impacts on the
possible formation of chemical gradients.

\begin{figure}
\begin{center}
\psfig{figure=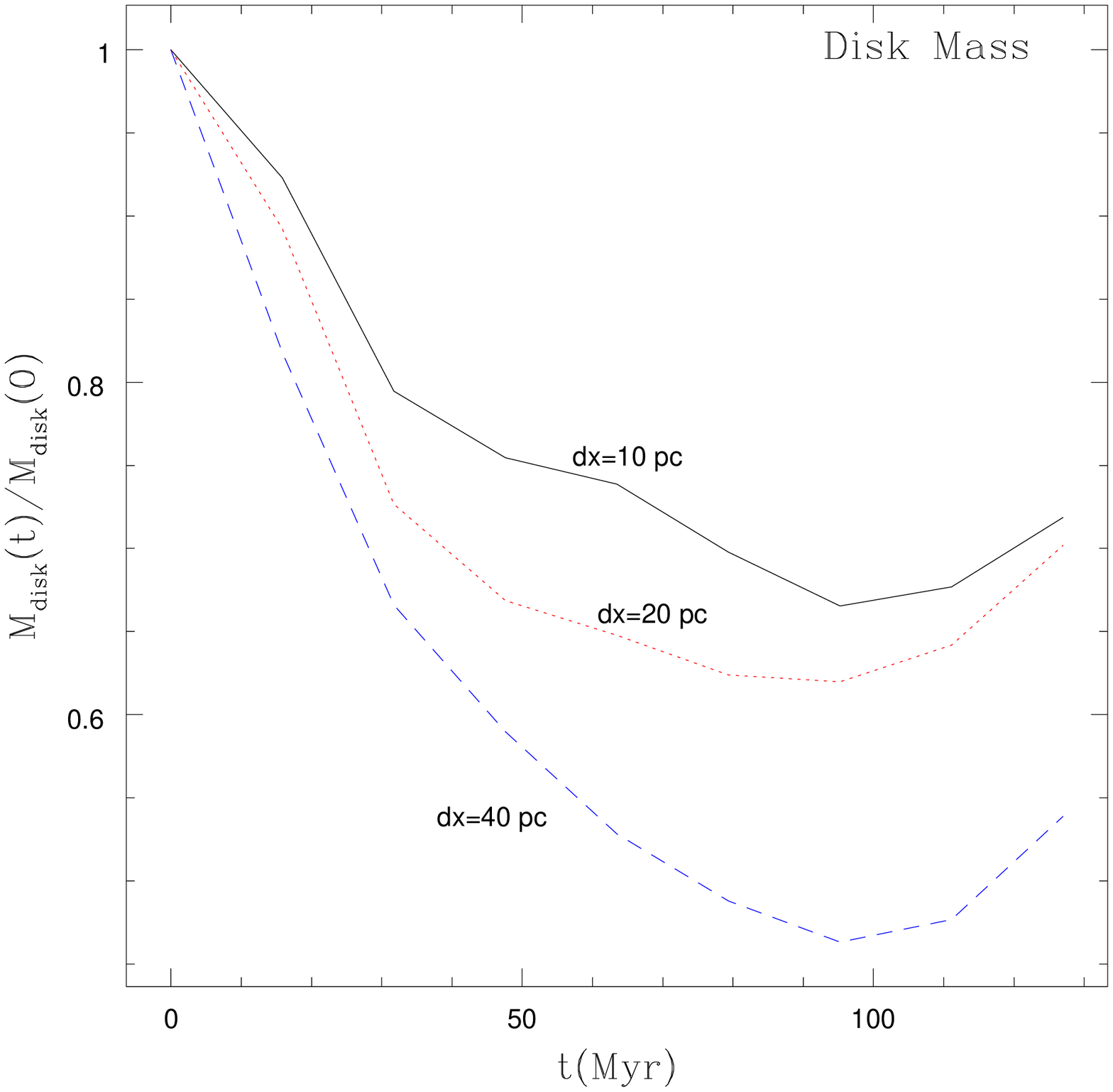,width=0.49\textwidth}
\end{center}
\caption{Time evolution of disk gas mass for different grid resolution for
our test model.
}
\label{fig:converg}
\end{figure}

To this aim, we run a test model with three different maximum
resolutions: $\Delta x={40,20,10}$ pc.  In order to save computational
time, the test model is a modified version of model MC3. Here the
computational grid covers the region $-2.5 < x < 2.5$ kpc, $-2.5 < y <
2.5$ kpc,  $z < 5$ kpc, while the
active region is limited to $R=1.5$ kpc. The results for the ISM mass
in the disk (one of the most sensitive quantity) are illustrated in
Figure \ref{fig:converg}, where it can be seen that $M_{\rm disk}(t)$ weakly
increases with the resolution. The reason for this trend is twofold:
$i$) the higher resolution generates smaller structures in the contact
discontinuities, increasing their total surface. As significant
radiative losses occur on these surfaces, less SN energy is available
to lift the ISM; $ii$) a higher resolution reduces the numerical
viscosity, diminishing the ability of the hot wind to drag the cold ISM.
As the models presented in the text are computed with $\Delta x=20$ pc
resolution, the values of the mass content found in the galactic disk
represent lower limits.

We examined also the behaviour of the ejecta mass in the disk. We find
that incresing the resolution to 10 pc its value increases by less
than 25\%, in line with the disk ISM trend. With a coarser grid with
zone size of 40 pc, however, results start to diverge more, with the
disk ejecta mass lowered by a factor of $\sim 4$.
We conclude that the resolution adopted  for our models (20 pc) is
high enough to give meaningful results.

\label{lastpage}
\end{document}